%% file: dis.tex
\documentclass[12pt]{report}
\usepackage[numbers,sort&compress]{natbib}
\usepackage{geometry}
\usepackage{fancyhdr}
\usepackage{afterpage}
\usepackage{graphicx}
\usepackage{amsmath,amssymb,amsbsy}
\usepackage{dcolumn,array}
\usepackage{tocloft}
\usepackage{asudis}
\usepackage[printonlyused]{acronym}
\usepackage{multirow}
\usepackage[space]{grffile}
\usepackage{xcolor}
\usepackage{morefloats}
\usepackage{footnote}
\usepackage{cleveref}
\usepackage{textcomp}
\usepackage{url}
\usepackage{caption}
\usepackage{subcaption}

\graphicspath{{Figures/}}

\let\ref\Cref
\begin{document}
\pagenumbering{roman}
\title{Computational Challenges in Non-parametric Prediction of Bradycardia in Preterm Infants}
\author{Sinjini Mitra}
\degreeName{Masters in Science (MS) in Electrical and Electronics Engineering}
\defensemonth{November}
\gradmonth{December}
\gradyear{2020}
\chair{Dr. Antonia Papandreou-Suppappola}
\cochair{Dr. Bahman Moraffah}
\memberOne{Dr. Pavan Turaga}
\maketitle
\doublespace
\include{abstract}
\acknowledgementpage{\input{ack.tex}}
\tableofcontents
\addtocontents{toc}{~\hfill Page\par}

\addcontentsline{toc}{part}{LIST OF FIGURES}
\renewcommand{\cftlabel}{Figure}
\listoffigures
\addtocontents{lof}{Figure~\hfill Page \par}
\newpage
\addcontentsline{toc}{part}{LIST OF TABLES}
\addtocontents{lot}{Table~\hfill Page \par}
\newpage
\addtocontents{toc}{CHAPTER \par}					
\listoftables


\doublespace
\pagenumbering{arabic}
\include{chapter1}

\include{chapter2}

\include{chapter3}

\include{chapter4}

\include{chapter5}
{\singlespace
\addcontentsline{toc}{part}{REFERENCES}
\bibliographystyle{IEEEtran}
\bibliography{references}}
\nocite{*}
\end{document}

%% file: abstract.tex
\begin{abstract}
Infants born before 37 weeks of pregnancy are considered to be preterm. Typically,  preterm infants have to be strictly monitored since they are highly susceptible to health problems like hypoxemia (low blood oxygen level), apnea, respiratory issues, cardiac problems, neurological problems as well as an increased chance of long-term health issues such as cerebral palsy, asthma and sudden infant death syndrome. One of the leading health complications in preterm infants is bradycardia -  which is defined as the slower than expected heart rate, generally beating lower than 60 beats per minute. Bradycardia is often accompanied by low oxygen levels and can cause additional long term health problems in the premature infant.

The implementation of a non-parametric method to predict the onset of bradycardia is presented. This method assumes no prior knowledge of the data and uses kernel density estimation to predict the future onset of bradycardia events. The data is preprocessed, and then analyzed to detect the peaks in the ECG signals, following which different kernels are implemented to estimate the shared underlying distribution of the data. The performance of the algorithm is evaluated using various metrics and the computational challenges and methods to overcome them are also discussed.

It is observed that the performance of the algorithm with regards to the kernels used are consistent with the theoretical performance of the kernel as presented in a previous work. The theoretical approach has also been automated in this work and the various implementation challenges have been addressed.
\end{abstract}

%% file: ack.tex
{This thesis would not have been possible without the support of
many people. I would like to thank my advisers Dr. Antonia Papandreou-Suppappola and Dr. Bahman Moraffah for their guidance and support in helping me finishing this thesis. Thanks to Dr. Pavan Turaga for agreeing to be on my defense committee. A special thank you to Dr. Thomas Matthew Holeva for his support and advice throughout the duration of my graduate education. Thanks to my dear friend Brian Baddadah and his many lessons on LaTeX. Thanks to my co-workers who put up with me while I rambled endlessly about this concept. Thanks to my parents Bajradeb Mitra and Laily Mitra who not only support me in every endeavor but have taught me to pursue my ambition with determination and hard work. This thesis would not be possible without their love. Thank you to my brothers Rhitabrata and Dhritabrata Mitra who encourage me and keep me going. Finally, thank you to my friends Nicole Martin, Madeline Damasco, Shayla Puryear and Andrew Wharton for their encouragement and support. }

%% file: chapter1.tex
\chapter{INTRODUCTION}
\label{chap:introduction}

\section{Motivation and existing methods}

Infants born before 37 weeks of pregnancy are considered to be preterm. Typically,  preterm infants have to be strictly monitored since they are highly susceptible to health problems like hypoxemia (low blood oxygen level), apnea, respiratory issues, cardiac problems, neurological problems as well as an increased chance of long-term health issues such as cerebral palsy, asthma and sudden infant death syndrome. One of the leading health complications in preterm infants is bradycardia -  which is defined as the slower than expected heart rate, generally beating lower than 60 beats per minute. Bradycardia is often accompanied by low oxygen levels and can cause additional long term health problems in the premature infant. Certain therapeutic interventions, as presented in \cite{thommandram2014rule, onak2019evaluation}, might be the most effective if intervention is done early in high-risk infants. However, in a study conducted with nineteen preterm infants (10 M/ 9 F) born between 25–33 weeks of gestation \cite{bib3},  it was found that even on the most sensitive setting of the oximeter, a significant number of bradycardias are not recorded.  Given the severity of the condition in terms of long term effects and the lack of a comprehensive and robust mechanism to predict bradycarida, there has been significant research and development in this area in recent times. \\

Existing methods as mentioned in \cite{williamson2013forecasting, lucchini2019multi}, use a combination of signal detection using multivariate feature construction from multimodal measurements, followed by machine learning to generate predictive warnings to aid real-time therapeutic interventions. The approach outlined in \cite{lucchini2019multi} uses Gaussian Mixture Models to successfully train the model and subsequently generate the the warnings. The limited success of these methods might help health care professionals and alert clinicians in the short term and ultimately provide automatic therapeutic care to reduce the complexity of predicting preterm cardiorespiratory conditions. However, there still remains a need to present a more robust and reliable method for the prediction of bradycardia.\\

This is addressed partly in \cite{gee2016predicting}, which hypothesizes that the immature cardiovascular control system in preterm infants exhibits transient temporal instabilities in heart rate that can be detected as a precursor signal of bradycardia. Statistical features in heartbeat signals prior to bradycardia are extracted and evaluated to assess their utility for predicting bradycardia. The  method uses point process analysis to generate real-time, stochastic measures from discrete observations of continuous biological mechanisms. This method achieves a false alarm rate of $0.79 \pm 0.018$. \\

Although the process in \cite{gee2016predicting} outperforms its predecessors, an alternate statistical method to predict the onset of bradycardia in preterm infants  is presented in \cite{bahman}. Unlike the parametric approach in \cite{gee2016predicting}, the method in \cite{bahman} assumes no prior knowledge of the data and uses non-parametric methods to predict the future onset of bradycardia events with 95\% accuracy. The data is modeled by first detecting the QRS complex in the ECG signals and then using kernel density estimator.  \\

\section{Proposed method}

The non-parametric estimation method for prediction of bradycardia mentioned in \cite{bahman} is implemented in this project through the creation of an automated algorithm to provide results consistent with the findings in the paper. The algorithm is created in R and MATLAB and uses non-parametric density estimation methods\cite{lantz} to predict the onset of bradycardia within the proposed failure rate of $5\%$. Although the theoretical method is discussed in detail \cite{bahman}, the implementation of said method is presented in this document. The aim of the document is to analyze the performance of the algorithm as well as to evaluate its performance based on common metrics. This would bridge the gap between the mathematical model and real-time application of the model. 

\section{Thesis organization}

This Thesis is organized as follows. Chapter 2 provides a brief background on non-parametric statistics and machine learning. In Chapter 3, we discuss a non-parametric density estimation based method to detect the bradycardia, introduce a hypothesis testing and the process used for algorithm automation. In Chapter 4, we present the main automation process and experimental results. We then conclude by discussing the future works in Chapter 5.

%% file: chapter2.tex
\chapter{BACKGROUND IN NON-PARAMETRIC STATISTICS AND MACHINE LEARNING}\label{chap:background}
\acresetall
This chapter seeks to provide an introductory background to the various statistical and machine learning concepts that have been used to automate the method presented in \cite{bahman}. Some of the key concepts covered are empirical risk minimization, bias variance decomposition, leave one out cross validation and non-parametric density estimation.\\

Statistics is the field of mathematics which involves the summary and analysis of numerical data in large quantities. The field of statistics can be divided into two general areas: descriptive statistics and inferential statistics. Descriptive statistics is a branch of statistics in which data are only used for descriptive purposes and are not employed to make predictions. Thus, descriptive statistics consists of methods and procedures for presenting and summarizing data. The procedures most commonly employed in descriptive statistics are the use of tables and graphs, and the computation of measures of central tendency and variability \cite{jaynes}.\\

Inferential statistics employs data in order to draw inferences (i.e., derive conclusions) or make predictions. Typically, in inferential statistics sample data are employed to draw inferences about one or more populations from which the samples have been derived.  Typically (although there are exceptions) the ideal sample to employ in research is a random sample. In a random sample, each subject or object in the population has an equal likelihood of being selected as a member of that sample. Machine learning draws heavily from statistics in not only classification and categorization of data but also for algorithm building and model selection\cite{jaynes}.

\section{Machine Learning methods} \label{sec: ml methods}

Machine learning can be broadly defined as computational methods using experience to improve performance or to make accurate predictions. Here, experience refers to the past information available to the learner, which typically takes the form of electronic data collected and made available for analysis. This data could be in the form of digitized human-labeled training sets, or other types of information obtained via interaction with the environment. In all cases, its quality and size are crucial to the success of the predictions made by the learner.
An example of a learning problem is how to use a finite sample of randomly selected documents, each labeled with a topic, to accurately predict the topic of unseen documents. Clearly, the larger is the sample, the easier is the task. But the difficulty of the task also depends on the quality of the labels assigned to the documents in the sample, since the labels may not be all correct, and on the number of possible topics \cite{tsybakov, mohri}.

Machine learning consists of designing efficient and accurate prediction algorithms. Some critical measures of the quality of these algorithms are their time and space complexity \cite{shalev, kuhn}.  However, in machine learning, there is a need for an additional notion of sample complexity to evaluate the sample size required for the algorithm to learn a family of concepts. 
Since the success of a learning algorithm depends on the data used, machine learning is inherently related to data analysis and statistics. More generally, learning techniques are data-driven methods combining fundamental concepts in computer science with ideas from statistics, probability and optimization.

Classification is the problem of assigning a category to each item in the data set. For example, document classification consists of assigning a category such as politics, business, sports, or whether to each document, while image classification consists of assigning to each image a category such as car, train, or plane. The number of categories in such tasks is often less than a few hundreds, but it can be much larger in some difficult tasks and even unbounded as in OCR, text classification, or speech recognition.

The different stages of machine learning are described below:
\begin{itemize}
\item $\emph{Examples}$: Items or instances of data used for learning or evaluation. This is also referred to as the data set.
\item $\emph{Features}$: The set of attributes, often represented as a vector, associated to an example.
\item $\emph{Labels}$: Values or categories assigned to examples. In classification problems, examples are assigned specific categories. In regression, items are assigned real-valued labels.
\item $\emph{Hyperparameters}$: Free parameters that are not determined by the learning algorithm, but rather specified as inputs to the learning algorithm.
\item $\emph{Training sample}$: Examples used to train a learning algorithm. The training sample varies for different learning scenarios as described in the next subsection.
\item $\emph{Validation samples}$: Examples used to tune the parameters of a learning algorithm when working with labeled data. The validation sample is used to select appropriate values for the learning algorithm's free parameters (hyperparameters).
\item $\emph{Test Sample}$: Examples used to evaluate the performance of a learning algorithm. The test sample is separate from the training and validation data and is not made available in the learning stage. The learning algorithm predicts labels for the test sample based on features. These predictions are then compared with the labels of the test sample to measure the performance of the algorithm.
\item $\emph{Loss function}$: A function that measures the difference, or loss, between a predicted label and a true label. Denoting the set of all labels as $y$ and the set of possible predictions as $y^{\prime}$ a loss function $L$ is a mapping $L: y \times y^{\prime} \rightarrow \mathbb{R}_{+}$. In most cases, $y^{\prime} = y$ and the loss function is bounded, but these conditions do not always hold.
\item $\emph{Hypothesis set}$: A set of functions mapping features (feature vectors) to the set of labels $y$. \\
\end{itemize}

\subsection{Learning scenarios}

The different learning scenarios for a learning algorithm differ in the types of training data available to the learner, the order and method by which training data is received and the test data used to evaluate the learning algorithm\cite{mohri, hastie}. 

\begin{itemize}
\item $\emph{Supervised learning}:$ The learner receives a set of labeled examples as training data and makes predictions for all unseen points. This is the most common scenario associated with classification, regression, and ranking problems.
\item $\emph{Unsupervised learning}:$ The learner exclusively receives unlabeled training data, and makes predictions for all unseen points. Since in general no labeled example is available in that setting, it can be difficult to quantitatively evaluate the performance of a learner. 
\item $\emph{Semi-supervised learning}:$ The learner receives a training sample consisting of both labeled and unlabeled data, and makes predictions for all unseen points. Semi-supervised learning is common in settings where unlabeled data is easily accessible but labels are expensive to obtain. Various types of problems arising in applications, including classification, regression, or ranking tasks, can be framed as instances of semi-supervised learning. The hope is that the distribution of unlabeled data accessible to the learner can help it achieve a better performance than in the supervised setting. The analysis of the conditions under which this can indeed be realized is the topic of much modern theoretical and applied machine learning research.
\item $\emph{Transductive inference}:$ As in the semi-supervised scenario, the learner receives a labeled training sample along with a set of unlabeled test points. However, the objective of transductive inference is to predict labels only for these particular test points. Transductive inference appears to be an easier task and matches the scenario encountered in a variety of modern applications. However, as in the semi-supervised setting, the assumptions under which a better performance can be achieved in this setting are research questions that have not been fully resolved.
\item $\emph{On-line learning}:$ In contrast with the previous scenarios, the online scenario involves multiple rounds where training and testing phases are intermixed. At each round, the learner receives an unlabeled training point, makes a prediction, receives the true label, and incurs a loss. The objective in the on-line setting is to minimize the cumulative loss over all rounds or to minimize the regret, that is the difference of the cumulative loss incurred and that of the best expert in hindsight. Unlike the previous settings just discussed, no distributional assumption is made in on-line learning. In fact, instances and their labels may be chosen adversarially within this scenario.
\item $\emph{Reinforcement learning}:$ The training and testing phases are also intermixed in reinforcement learning. To collect information, the learner actively interacts with the environment and in some cases affects the environment, and receives an im- mediate reward for each action. The object of the learner is to maximize his reward over a course of actions and iterations with the environment. However, no long-term reward feedback is provided by the environment, and the learner is faced with the exploration versus exploitation dilemma, since he must choose between exploring unknown actions to gain more information versus exploiting the information already collected.
\item $\emph{Active learning}:$ The learner adaptively or interactively collects training examples, typically by querying an oracle to request labels for new points. The goal in active learning is to achieve a performance comparable to the standard supervised learning scenario (or passive learning scenario), but with fewer labeled examples. Active learning is often used in applications where labels are expensive to obtain, for example computational biology applications.
\end{itemize}

\subsection{Model Selection}\label{model}

A key problem in learning algorithm is the selection of the hypothesis set $\mathcal{H}$ . The choice of $\mathcal{H}$ is subject to a trade-off that can be analyze using estimation and approximation errors\cite{mohri, shalev}. 

Let $\mathcal{H}$ be a family of functions mapping $\mathcal{X}$ to $\{-1,+1\}$. The excess error of a hypothesis $h$ chosen from $\mathcal{H}$, that is the difference between its error $R(h)$ and the Bayes error $R^{*}$, can be decomposed as follows:

$$R(h)-R^{*}=\underbrace{\left(R(h)-\inf _{h \in \mathcal{H}} R(h)\right)}_{\text {estimation }}+\underbrace{\left(\inf _{h \in \mathcal{H}} R(h)-R^{*}\right)}_{\text {approximation }}$$

The first term is called the estimation error, the second term the approximation error. The estimation error depends on the hypothesis $h$ selected. It measures the error of $h$ with respect to the infimum of the errors achieved by hypotheses in $\mathcal{H}$. 
The approximation error measures how well the Bayes error can be approximated using $\mathcal{H}$. It is a property of the hypothesis set $\mathcal{H}$, a measure of its richness. For a more complex or richer hypothesis $\mathcal{H}$, the approximation error tends to be smaller at the price of a larger estimation error. Model selection consists of choosing $\mathcal{H}$ with a favorable trade-off between the approximation and estimation errors.

\subsubsection{Empirical Risk Minimization}\label{emp}

Empirical Risk Minimization is a standard algorithm by which the estimation error can be bounded. ERM seeks to minimize the error on the training sample and the following discussion is adapted from \cite{mohri}:

$$h_{S}^{\mathrm{ERM}}=\underset{h \in \mathcal{H}}{\operatorname{argmin}} \widehat{R}_{S}(h)$$

where $\widehat{R}_{S}(h)$ is the calculated error on the training sample composed of data 

\begin{figure}[t!]
\begin{center}
\includegraphics[width= 8cm, height = 6cm]{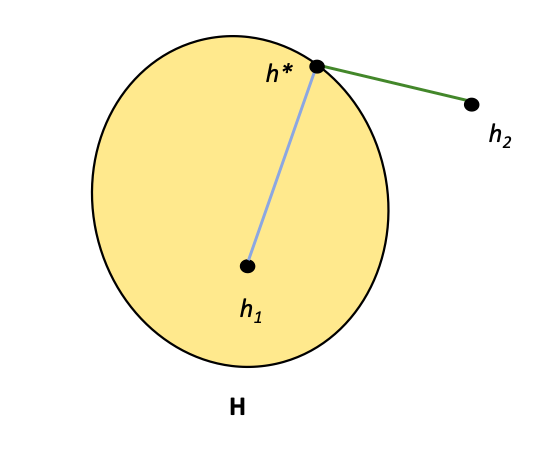}
\caption{The estimation and approximation error are represented in blue and green respectively. These errors are the distance of a chosen hypothesis $h^{*}$  from $h_{1}$ and $h_{2}$ }
\label{fig: herror}
\end{center}
\end{figure}

derived from the sample space $S$. \\

Let us consider that there is a joint probability distribution $P(x,y)$ over $X$ and $Y$, and the training set consists of $n$ instances $\left(x_{1}, y_{1}\right), \ldots,\left(x_{n}, y_{n}\right)$ drawn i.i.d. from $P(x,y)$. It is also assumed there is a non-negative real-valued loss function $L(\hat{y}, y)$ which measures how different the prediction $\hat {y}$ of a hypothesis is from the true outcome $y$. The hypothesis $h$ is a function of $x$, therefore, $\hat{y} = h(x)$. The loss function can be written as $L(\hat{y},y) = L(h(x),y) = h(x)-y$\cite{mohri}. 
The risk associated with hypothesis $h(x)$ is then defined as the expectation of the loss function $L(h(x),y)$:

$$R(h)=\mathbf{E}[L(h(x), y)]=\int L(h(x), y)  d P(x, y)$$

In general, the risk $R(h)$ cannot be computed because the distribution $P(x,y)$ is unknown to the learning algorithm. However, an approximation (called empirical risk) is calculated by averaging the loss function on the training set:

$$R_{\mathrm{emp}}(h)=\frac{1}{n} \sum_{i=1}^{n} L\left(h\left(x_{i}\right), y_{i}\right),$$

where $x_{i}$ and $y_{i}$ are data points drawn from the training set of length $n$. Thus the learning algorithm defined by the ERM principle consists in solving the above optimization problem \cite{mohri}, \cite{hastie}. The concept of the loss function $L(h(x),y)$ and risk  is elaborated further in \ref{mse} and \ref{cv}.

\section{Bias Variance Decomposition} \label{bvd}

The loss function in  \ref{emp} is the $L_{2}$ loss  which is calculated using the calculated estimate and the true value of the estimate at a particular point. For any such estimation, there are a number of errors present. This section aims to analyze such errors and present a bound on the $L_{2}$ loss as observed in \cite{hastie}. Given a data set $D = (x_{1}, y_{1}), (x_{2}, y_{2}), . . ., (x_{n},y_{n})$ it is assumed these points are independent identically distributed (i.i.d) and drawn from the same unknown distribution $P(x,y)$. For non-parametric density estimation the parameters of $P(x,y)$ are unknown but it is still possible to estimate $P(x,y)$ and this estimate is denoted by $f$. If this $f$ is used to design a hypothesis to classify our given data set according to some condition, there appears two sources of errors in the estimation of $f$ as elaborated in \cite{hastie}:

1. $\textbf{Bias}$: This kind of error is caused by the inability of $f$ to estimate $P(x,y)$ correctly. 

2. $\textbf{Variance}$: This kind of error is caused by the presence of random noise in the data which causes the estimate $f$ to be inaccurate and vary from $P(x,y)$

Let us assume any arbitrary point $z$ drawn from the distribution $P(x,y)$ such that it is not a sample data point in the dataset $D$. The estimate at $z$ is represented by 
$$\gamma = f(z)+\epsilon.$$

Note that $\gamma$ is random. It is worth mentioning $f(z)$ is the value of the estimate of $f$ at $z$ and $\epsilon$ is added noise such that 
\begin{flalign}
\begin{split}
\mathrm{E}[\gamma] &= f(z),\\
\text{Var}(\gamma) &= \text{Var}(\epsilon)
\end{split}
\end{flalign}

A square loss is assumed, $|P(z) - f(z)|^2$, the difference between the value of the actual distribution $P$ and the estimate $f$ at $z$. The risk which is the expectation of the squared loss is computed. Risk, $R(z)$ is written as 

\begin{figure}[t!]
\begin{center}
\includegraphics[width= 12cm, height=10 cm]{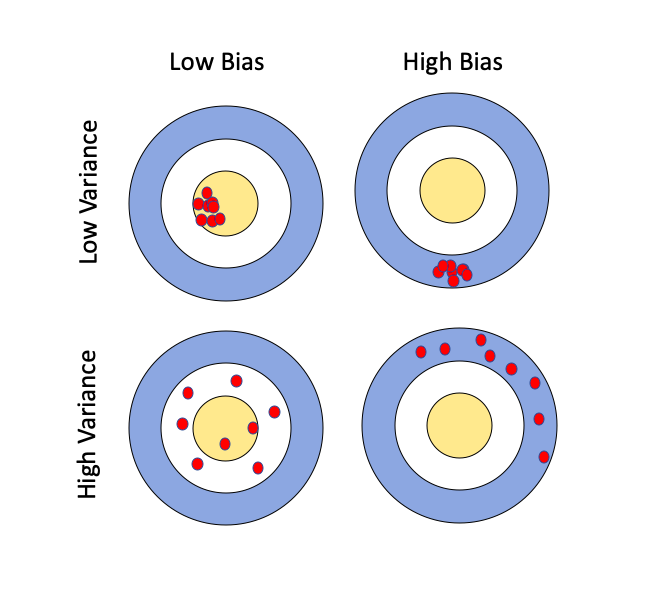}
\caption{ Illustration of the presence of bias/variance (or both) in a data set. }
\label{fig:varmatrix }
\end{center}
\end{figure} 

\begin{flalign}
\begin{split}
R(z)&= \mathrm{E}\left[(P(z)-\gamma)^{2}\right]\\
&=\mathrm{E}\left[P(z)^{2}\right]+\mathrm{E}\left[\gamma^{2}\right]-2 \mathrm{E}[\gamma P(z)]\\
&=\operatorname{Var}(P(z))+\mathrm{E}[P(z)]^{2}+\operatorname{Var}(\gamma)+\mathrm{E}[\gamma]^{2}-2 \mathrm{E}[\gamma] \mathrm{E}[P(z)]\\
&=(\mathrm{E}[P(z)]-\mathrm{E}[\gamma])^{2}+\operatorname{Var}(P(z))+\operatorname{Var}(\gamma)\\
&=(\mathrm{E}[P(z)]-f(z))^{2}+ \operatorname{Var}(P(z))+ \operatorname{Var}(\epsilon)
\end{split}
\end{flalign}

$\mathrm{E}[P(z)]-f(z)$ is defined to be the bias. If the bias is larger than zero,the estimator is said to be positively biased, if the bias is smaller than zero, the estimator is negatively biased, and if the bias is exactly zero, the estimator is unbiased. The higher the bias, the worse our estimate is a fit for the actual distribution $P(x,y)$. Bias is inherent to a particular model and a large amount of bias suggests underfitting \cite{tsybakov, mohri}. 

$\operatorname{Var}(P(z))$ is the variance of the estimator (kernel density, regression, etc.). The variance as the difference between the expected value of the squared estimator minus the squared expectation of the estimator. A large variance may indicate that our model is highly specialized to only one dataset, model is extremely complicated,  which indicates overfitting. 

 $\operatorname{Var}(\epsilon)$ is called irreducible error. It is an error that cannot be decomposed further and will always be present in the estimate. This is also called noise. 

\begin{figure}
\includegraphics[width=\linewidth]{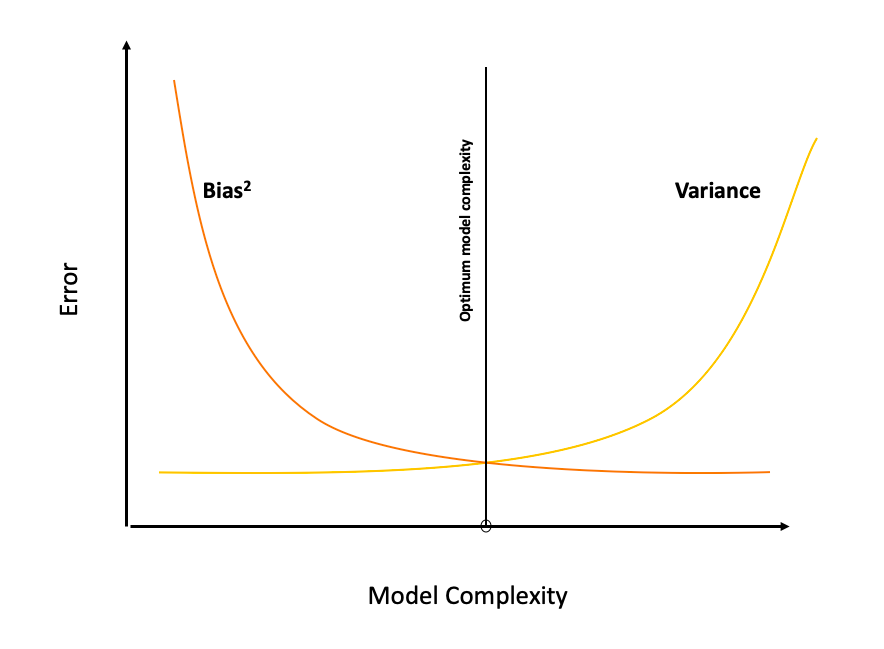}
\caption{Figure illustrating the bias variance decomposition and the optimal complexity}
\label{fig: decomp1}
\end{figure}

In real world scenarios, $P(x,y)$ is rarely known and especially in the case of non-parametric estimations, we lack knowledge about the parameters of $P(x,y)$. However, in calculating the estimate at arbitrary points, as seen above, there exists three different sources of error. Therefore, it is necessary to minimize the risk of any estimation method so that the bias and variance are both minimized. There is however, a trade-off between the two \cite{hastie, shalev}. 

As seen in \ref{fig: decomp1}, the most complex models overfit the data while the simple models underfit the data. Therefore the optimum complexity for a model is given by the values of bias and variance which minimize the total error.

\section{ Density Estimation}\label{estimation}

Density estimation is the process of reconstructing the probability density using a set of given data points or observations. In statistics, a random variable is a variable whose value depends on a random phenomenon \cite{murphy}. For instance, the outcome of a coin toss, the event that it rains in Phoenix tomorrow or the sum of any two numbers between 0 and 1 are all random variables. A random variable can be either discrete (having specific values) or continuous (any value in a continuous range). Some outcomes of a random variable are more likely to occur (high probability density) and other outcomes are less likely to occur (low probability density). 
The overall shape of the probability density is referred to as a probability distribution, and the calculation of probabilities for specific outcomes of a random variable is performed by a probability density function (PDF).

\begin{figure}[t!]
\begin{center}
\includegraphics[width= \linewidth]{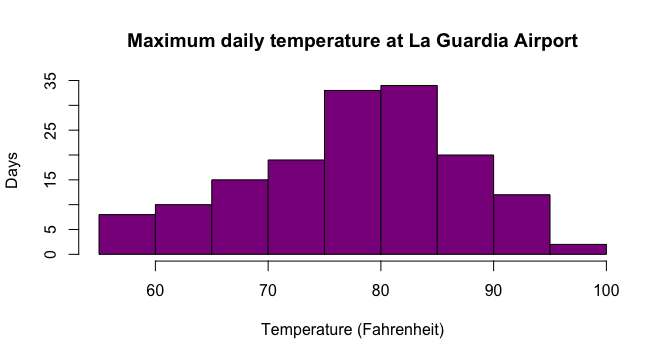}
\caption{ A histogram showing the maximum daily temperatures at La Guardia Airpot, New York. The data is drawn from the $\emph{airquality}$ dataset in R which contains the daily air quality measurements in New York, May-September, 1973}
\label{fig:hist }
\end{center}
\end{figure}

From the PDF of certain data it can be judged whether a given observation is likely, or unlikely. One of the most popular and common ways to visualize the probability density function is the histogram (as seen in \ref{fig:hist }). A histogram is a graph of the frequency distribution in which the vertical axis represents the count (frequency) and the horizontal axis represents the possible range of the data values. A histogram is created by dividing up the range of the data into a small number of intervals or bins. The number of observations falling in each interval is counted. This gives a frequency distribution \cite{freedman}. 

The shape of the probability density function across the domain for a random variable is referred to as the probability distribution and common probability distributions have names, such as uniform, normal, exponential, and so on. Given a random variable the aim is the estimation of a density of its probabilities. Therefore, it follows that one would want to know what the probability density looks like. However, in most cases the distribution of the random variable is not known. This is because there is no available knowledge of all possible outcomes for a random variable, only a small set of observations. However, the distribution can be estimated and this is called density estimation. In density estimation, the small sample of observed variables are used to estimate the overall probability distribution \cite{lantz}.

Generalizing further to the case of any distribution of a set of random variables, let  $D = X_{1},X_{2},...., X_{n}$ be independent identically distributed (i.i.d.) real valued random variables that share a common distribution. The density of this distribution, denoted by $p(x)$, is a function on $\mathbb{R}$ from $[0,+\infty)$, but is unknown. An estimator of $p(x)$ is a function $x \mapsto p_{n}(x)=p_{n}\left(x, X_{1}, \ldots, X_{n}\right)$ measurable with respect to the observations $\mathbf{X}=\left(X_{1}, \ldots, X_{n}\right)$. 

As described earlier, inferential statistics is aimed at making inferences about the larger population from which the samples are drawn. The main goals of inferential statistics are: parameter estimation, data prediction and model comparison. There are to main approaches in inferential statistics:

\begin{itemize}
\item
$\emph{Frequentist}$: The frequentist school only uses conditional distributions of data given specific hypotheses. The presumption is that some hypothesis (parameter specifying the conditional distribution of the data) is true and that the observed data is sampled from that distribution. In particular, the frequentist approach does not depend on a subjective prior that may vary from one investigator to another. In frequentist approaches, only repeatable random events have probabilities. These probabilities are equal to the long-term frequency of occurrence of the events in question. No probability is attached to hypotheses or to any fixed but unknown values \cite{silverman1986density}.
\item
$\emph{Bayesian}$: In contrast, the Bayesian school models uncertainty by a probability distribution over hypotheses.  The ability to make inferences depends on the degree of confidence in the chosen prior, and the robustness of the findings to alternate prior distributions may be relevant and important. Bayesian approaches associate probabilities to any event or hypotheses. Probabilities are also attached to non-repeatable events\cite{murphy}. 
\end{itemize}

Frequentist measures like $p$-values and confidence intervals continue to dominate research,
especially in the life sciences. However, in the current era of powerful computers and
big data, Bayesian methods have undergone an enormous renaissance in fields like machine learning and genetics. In this method a Bayesian approach is used to estimate the unknown density function.

\subsection{Parametric Density Estimation} \label{pdest}

If it is known a priori that $p(x)$ belongs to a parametric family $\{g(x, \theta): \theta \in \Theta\}$, where $g(\cdot, \cdot)$ is a given function, and $\Theta $ is a subset of $\mathbb{R}^{k}$ with a fixed dimension $k$ independent of $n$, the estimator for $p(x)$ is equivalent to the estimation of the finite-dimensional parameter $\theta$ . This is a parametric problem of estimation \cite{mohri, murphy}. 

If the shape of the unknown distribution follows well-known sets/classes of distributions it can be estimated using certain parameters (like the mean, median, standard deviation etc.). For instance, the normal distribution has two parameters: the mean and the standard deviation. Given these two parameters, the probability distribution function is now known. These parameters can be estimated from data by calculating the sample mean and sample standard deviation. This process is called parametric density estimation \cite{gelman}. There are two ways in which the parameter can be estimated:\\

$\emph{Maximum Likelihood Estimation}:$ In this case, the parameter $\theta$ is unknown but fixed. Given the data, $\theta$ is chosen such that it maximizes the probability of obtaining the samples that have already been observed.the following discussion has been adapted from \cite{gelman}. The density $p(x)$ is completely specified by parameter $\theta = [\theta_{1}, \theta_{2}, \dots,\theta_{k}]$ . If $p(x)$ is Gaussian with $N\left(\mu, \sigma^{2}\right)$ then $\theta=\left[\mu, \sigma^{2}\right]$. Since $p(x)$ depends on $\theta$, it can be denoted by $p(x \mid \theta)$, where $p(x \mid \theta)$ is not a conditional density but only demonstrates dependence. If $p(x)$ is $N\left(\mu, \sigma^{2}\right)$ then $D = X_{1},X_{2},...., X_{n}$ are i.i.d. samples from $N\left(\mu, \sigma^{2}\right)$ and,

\begin{equation}\label{eq:independent}
p(D \mid \theta)=\prod_{k=1}^{k=n} p\left(x_{k} \mid \theta\right)=F(\theta)
\end{equation}

\ref{eq:independent} is called the likelihood of $\theta$ with respect to the observations $D$. The value of $\theta$ that maximizes the likelihood function $p(D \mid \theta)$ is given by 

\begin{equation}\label{eq:argmax}
\hat{\theta}=\underset{\theta}{\arg \max }(p(D \mid \theta))
\end{equation}

Instead of maximizing $p(D \mid \theta)$ it is often easier to maximize $\operatorname{In}(p(D \mid \theta))$. Since log is a monotonic function \ref{eq:argmax} is rewritten as, 

$$\underset{\theta}{\arg \max }(p(D \mid \theta)) = \underset{\theta}{\operatorname{argmax}}(\operatorname{In} p(D \mid \theta))$$

Therefore,

\begin{equation} \label{eq:1}
\hat{\theta} = \underset{\theta}{\operatorname{argmax}}(\operatorname{In} p(D \mid \theta)) =\underset{\theta}{\arg \max }\left(\ln \prod_{k=1}^{k=n} p\left(x_{k} \mid \theta\right)\right)
= \underset{\theta}{\arg \max }\left(\sum_{k=1}^{n} \ln p\left(x_{k} \mid \theta\right)\right)
\end{equation}

Let us consider the Gaussian parameter mentioned before. It is assumed that $p(x \mid \mu)$ is $N\left(\mu, \sigma^{2}\right)$ where $\sigma^{2}$ is known but $\mu$ is unknown and needs to be estimated. Therefore, $\theta = \mu$ for this problem and using \ref{eq:1} it can be concluded that:

$$\begin{aligned}\hat{\mu} =  &\underset{\mu}{\arg \max }\left(\sum_{k=1}^{n} \ln p\left(x_{k} \mid \mu\right)\right)\\
&=\underset{\mu}{\arg \max }\left(\sum_{k=1}^{n} \ln \left(\frac{1}{\sqrt{2 \pi \sigma}} \exp \left(-\frac{\left(x_{k}-\mu\right)^{2}}{2 \sigma^{2}}\right)\right)\right)\\
&=\underset{\mu}{\arg \max } \sum_{k=1}^{n}\left(-\ln \sqrt{2 \pi \sigma}-\frac{\left(x_{k}-\mu\right)^{2}}{2 \sigma^{2}}\right)\\
\end{aligned}$$

For easier notation, the previous equation is rewritten as $M(\mu)$, and subsequently,

\begin{equation}\label{eq:2}
\underset{\mu}{\arg \max }(M(\mu))=\underset{\mu}{\arg \max } \sum_{k=1}^{n}\left(-\ln \sqrt{2 \pi \sigma}-\frac{\left(x_{k}-\mu\right)^{2}}{2 \sigma^{2}}\right)
\end{equation}

Taking the derivative of \ref{eq:2},

$$\frac{d}{d \mu}(M(\mu))=\sum_{k=1}^{n} \frac{1}{\sigma^{2}}\left(x_{k}-\mu\right)=0$$

Simplifying further,

$$\sum_{k=1}^{n} x_{k}-n \mu=0$$ and finally,

\begin{equation}\label{eq:3}
\hat{\mu}=\frac{1}{n} \sum_{k=1}^{n} x_{k}
\end{equation}

As seen in \ref{eq:3}, the maximum likelihood estimator of the mean is just the average value of the observed samples, $D$. \ref{eq:3} makes intuitive sense since in general, one would assume that the mean of a set of data is the numerical average.

$\emph{Bayesian Estimation}:$ In this method, the observed data is fixed and different values of $\theta$ are assumed. Therefore, unlike the maximum likelihood approach, $\theta$ is now the random variable.  The following discussion has been adapted from \cite{manning}. The aim is the estimation of $\theta$ given $D= X_{1},X_{2},...., X_{n}$. It is assumed that $\theta$ is continuous. The posterior probability distribution of $\theta$ is given by $p(\theta \mid I)$ and it should be normalized such that

\begin{equation} \label{eq:posterior}
\int_{-\infty}^{\infty} p(\theta \mid I) d \theta=1
\end{equation}

$p(\mathbf{D} \mid \theta, I)$ is the sampling distribution for $D$ given the model implied by $I$ and $\theta$. Baye's Theorem tells us that posterior distribution for $\theta$ should be

\begin{equation}\label{eq:bayes}
p(\theta \mid \mathbf{D}, I)=\frac{p(\mathbf{D} \mid \theta, I) p(\theta \mid I)}{p(\mathbf{D} \mid I)}
\end{equation}

It is true that \ref{eq:bayes} is also normalized such that, 

$$\int_{-\infty}^{\infty} p(\theta \mid \mathbf{D}, I) d \theta=1$$

The denominator $p(\mathbf{D} \mid I)$ in \ref{eq:bayes} can be evaluated as

\begin{equation}\label{eq:dataprob}
p(\mathbf{D} \mid I)=\int_{-\infty}^{\infty} p(\mathbf{D}, \theta \mid I) d \theta=\int_{-\infty}^{\infty} p(\mathbf{D} \mid \theta, I) p(\theta \mid I) d \theta
\end{equation}

Since \ref{eq:dataprob} only depends on $D$ and not the parameter $\theta$, one can say that \ref{eq:bayes} is actually

$$p(\theta \mid \mathbf{D}, I) \propto p(\mathbf{D} \mid \theta, I) p(\theta \mid I)$$

where  $\propto$ implies that there is some numerical constant $M$ that equates the terms on the left of $\propto$ to the terms on the right. The value of $M$ does not depend on $\theta$ but may depend on $D$. 

$$p(\theta \mid \mathbf{D}, I)=M(\mathbf{D}) p(\mathbf{D} \mid \theta, I) p(\theta \mid I)$$

$M$ can be calculated using 

$$\int_{-\infty}^{\infty} p(\theta \mid \mathbf{y}, I) d \theta=1$$

The Gaussian case is considered where only $\mu$ is unknown $(\theta = \mu)$. Therefore, there is a need to establish the posterior probability of $\mu$ denoted by $p(\mu)$. Assuming that

\begin{equation} \label{eq:6}
p(x \mid \mu) \sim \mathrm{N}\left(\mu, \sigma^{2}\right),
\end{equation}

and,

\begin{equation} \label{eq:eq7}
p(\mu) \sim \mathrm{N}\left(\mu_{0}, \sigma_{0}^{2}\right),
\end{equation}
where $\sigma, \mu_{0}, \sigma_{0}$ are known to us.

Here, $p(\mu)$ is akin to $p(\theta \mid I)$ (the posterior probability of $\theta$). The aim is to find $p(\theta \mid \mathbf{D}, I)$ using \ref{eq:bayes}. By assumption of independence,

\begin{equation} \label{eq:indbayes}
p(\mathbf{D} \mid \theta, I)=\prod_{k=1}^{k=n} p\left(x_{k} \mid \theta\right)
\end{equation}

To solve for \ref{eq:bayes}, \ref{eq:6}, \ref{eq:eq7} and \ref{eq:indbayes} are used.

\begin{equation}\label{eq:eq8}
p(\mu \mid \mathbf{D}, I)=\alpha \prod_{k=1}^{n} p\left(x_{k} \mid \mu\right) p(\mu)
\end{equation}
where $\alpha$ is the scale parameter as discussed previously and is independent of $\mu$. As $x_{k}$ is normally distributed, $p\left(x_{k} \mid \mu\right)$ and $p(\mu)$ are updated with the specific equations:

\begin{equation}\label{eq:eq9}
p\left(x_{k} \mid \mu\right)=\frac{1}{\left(2 \pi \sigma^{2}\right)^{1 / 2}} \exp \left[-\frac{1}{2}\left(\frac{x_{k}-\mu}{\sigma}\right)^{2}\right]
\end{equation}

\begin{equation}\label{eq:eq10}
p(\mu)=\frac{1}{\left(2 \pi \sigma_{0}^{2}\right)^{1 / 2}} \exp \left[-\frac{1}{2}\left(\frac{\mu-\mu_{0}}{\sigma_{0}}\right)^{2}\right]
\end{equation}

Substituting \ref{eq:eq9} and \ref{eq:eq10} in \ref{eq:eq8},

$$p(\mu \mid \mathbf{D})=\alpha \prod_{k=1}^{n} \frac{1}{\left(2 \pi \sigma^{2}\right)^{1 / 2}} \exp \left[-\frac{1}{2}\left(\frac{x_{k}-\mu}{\sigma}\right)^{2}\right] \frac{1}{\left(2 \pi \sigma_{0}^{2}\right)^{1 / 2}} \exp \left[-\frac{1}{2}\left(\frac{\mu-\mu_{0}}{\sigma_{0}}\right)^{2}\right]$$

$$p(\mu \mid \mathbf{D})=\alpha \prod_{k=1}^{n} \frac{1}{\left(2 \pi \sigma^{2}\right)^{1 / 2}} \frac{1}{\left(2 \pi \sigma_{0}^{2}\right)^{1 / 2}} \exp \left[-\frac{1}{2}\left(\frac{\mu-\mu_{0}}{\sigma_{0}}\right)^{2}-\frac{1}{2}\left(\frac{x_{k}-\mu}{\sigma}\right)^{2}\right]$$

Simplifying further,

$$p(\mu \mid \mathbf{D})=\alpha^{\prime} \exp \sum_{k=1}^{n}\left(-\frac{1}{2}\left(\frac{\mu-\mu_{0}}{\sigma_{0}}\right)^{2}-\frac{1}{2}\left(\frac{x_{k}-\mu}{\sigma}\right)^{2}\right)$$

where $$\alpha^{\prime} = \frac{1}{\left(2 \pi \sigma^{2}\right)^{1 / 2}} \frac{1}{\left(2 \pi \sigma_{0}^{2}\right)^{1 / 2}}$$

On further simplification,

\begin{equation}\label{eq:eq11}
p(\mu \mid \mathbf{D})=\alpha^{\prime \prime} \exp \left[-\frac{1}{2}\left(\frac{n}{\sigma^{2}}+\frac{1}{\sigma_{0}^{2}}\right) \mu^{2}-2\left(\frac{1}{\sigma^{2}} \sum_{k=1}^{n} x_{k}+\frac{\mu_{0}}{\sigma_{0}^{2}}\right) \mu\right]
\end{equation}

Comparing \ref{eq:eq11} to the Gaussian distribution in the standard form:

\begin{equation}\label{eq:post}
p(\mu \mid \mathbf{D})=\frac{1}{\left(2 \pi \sigma_{n}^{2}\right)^{1 / 2}} \exp \left[-\frac{1}{2}\left(\frac{\mu-\mu_{n}}{\sigma_{n}}\right)^{2}\right]
\end{equation}

where,

\begin{equation}\label{eq:mun}
\mu_{n}=\left(\frac{n \sigma_{0}^{2}}{n \sigma_{0}^{2}+\sigma^{2}}\right) \overline{x_{n}}+\frac{\sigma^{2}}{n \sigma_{0}^{2}+\sigma^{2}} \mu_{0}
\end{equation}

and,

\begin{equation}\label{eq:sigman}
\sigma_{n}^{2}=\frac{\sigma_{0}^{2} \sigma^{2}}{n \sigma_{0}^{2}+\sigma^{2}}
\end{equation}

It is important to note that for Gaussian random variables the variance and the mean are the required parameters to estimate the underlying distribution. Therefore, it is important to emphasize both $\mu_{n}$ and $\sigma_{n}^{2}$.\\

In order to find $p(x \mid \mathbf{D})$ the following equation is used

\begin{equation}\label{eq:eq13}
p(x \mid \mathbf{D})=\int p(x, \theta \mid \mathbf{D}) d \theta=\int p(x \mid \theta) p(\theta \mid \mathbf{D}) d \theta
\end{equation}

$p(\mu \mid \mathbf{D})$ is given by \ref{eq:post} and $p(x \mid \mu)$ is given by \ref{eq:6}. Therefore, \ref{eq:eq13} can be rewritten as:

$$p(x \mid \mathbf{D})=\int p(x \mid \mu) p(\mu \mid \mathbf{D}) d \mu$$

$$p(x \mid \mathbf{D})=\int \frac{1}{\sqrt{2 \pi} \sigma} \exp \left[-\frac{1}{2}\left(\frac{x-\mu}{\sigma}\right)^{2}\right] \frac{1}{\sqrt{2 \pi} \sigma_{n}} \exp \left[-\frac{1}{2}\left(\frac{\mu-\mu_{n}}{\sigma_{n}}\right)^{2}\right] d \mu$$

Substituting the \ref{eq:mun} and \ref{eq:sigman},

$$p(x \mid \mathbf{D})=\frac{1}{2 \pi \sigma \sigma_{n}} \exp \left[-\frac{1}{2} \frac{(x-\mu)}{\sigma^{2}+\sigma_{n}^{2}}\right] \int \exp \left[-\frac{1}{2} \frac{\sigma^{2}+\sigma_{n}^{2}}{\sigma^{2} \sigma_{n}^{2}}\left(\mu-\frac{\sigma_{n}^{2} \bar{x}_{n}+\sigma^{2} \mu_{n}}{\sigma^{2}+\sigma_{n}^{2}}\right)^{2}\right] d \mu$$

Therefore, $p(x \mid \mathbf{D})$ is distributed normally as,

$$p(x \mid D) \sim N\left(\mu_{n}, \sigma^{2}+\sigma_{n}^{2}\right)$$

Thus, it is observed that for Bayesian parametric estimation, the underlying distribution is estimated using the posterior probability and conditional densities obtained from the given data.


\subsection{Non-Parametric Density Estimation}\label{npd}

In some cases however, a priori knowledge about $p(x)$ is not known. In this case it is usually assumed that $p(x)$ belongs to some class $\mathcal{P}$ of densities \cite{mohri, tsybakov}. For example, $\mathcal{P}$ can be the set of all the continuous probability densities on $\mathbb{R}$. In this case, parametric density estimation is not feasible and alternative methods must be used to estimate the density $p(x)$. The distributions still have parameters, but they cannot be controlled or used in the process of estimation since they are unknown. One of the most common non-parametric density approaches is the kernel density estimator (KDE) \cite{tsybakov} which is elaborated in the next section. 

\subsection{Kernel Density Estimator}\label{kde}

A kernel is a mathematical weighting function that returns a probability for a given value of a random variable. The kernel effectively smooths or interpolates the probabilities across the range of outcomes for a random variable such that the sum of probabilities equals one \cite{jaynes}.\\

$\emph{Smoothing parameter (h)}:$ The smoothing parameter or bandwidth controls how wide the probability mass is spread around a particular point as well as controlling the smoothness or roughness of a density estimate. In other words, the bandwidth controls the number of samples or window of samples used to estimate the probability for a new point. It is often denoted by $h$. It is important to note that this $h$ is different from the hypothesis $h$ mentioned in \ref{model} and henceforth the use of $h$ in this document denotes the bandwidth of a kernel. A large bandwidth may result in a rough density with few details, whereas a small window may have too much detail but not be general enough to cover any new unseen samples. Therefore, it is important to select the correct bandwidth $h$. In this project, the best value of $h$ is learned through leave one out cross validation\cite{Ala14} elaborated in \ref{cv}. 

The contribution of samples in a window is governed by one of the many kernel basis functions denoted by $K(\cdot)$. A list of the kernel basis functions used in this project can be found in \ref{table:kernel}.

The derivation for the kernel density estimator mentioned in the following discussion has been adapted from \cite{tsybakov}. Generalizing to the case of a set of random variables $D={X_{1}, X_{2}, X_{3},....,X_{n}}$ which are independent identically distributed (i.i.d) and have a probability density $p(x)$, for any small $h > 0$, $p(x)$ can be calculated as

\begin{equation}\label{eq:kern1}
p(x) \approx \frac{F(x+h)-F(x-h)}{2 h},
\end{equation}

 such that $x \in \mathcal{R}$ and $F(x)$ is defined as

$$F_{n}(x)=\frac{1}{n} \sum_{i=1}^{n} I\left(X_{i} \leq x\right)$$

where $I()$ is the indicator function. By the strong law of large numbers,

$$F_{n}(x) \rightarrow F(x), \quad \forall x \in \mathcal{R}$$

which allows us to replace $F(x)$ with $F_{n}(x)$ in \ref{eq:kern1},

\begin{equation}\label{eq:rosen}
\hat{p}_{n}^{R}(x)=\frac{F_{n}(x+h)-F_{n}(x-h)}{2 h}
\end{equation}

\begin{table}[t!]
\begin{center}
\begin{tabular}{| c | c |}
\hline
Kernel & Equation \\ 
\hline\hline
Gaussian 
& 
$K(u)=\frac{1}{\sqrt{2 \pi}} e^{-\frac{1}{2} u^{2}}$\\
\hline
Epanechnikov (parabolic)
&
$K(u)=\frac{3}{4}\left(1-u^{2}\right)$
, support: $|u| \leq 1$\\
\hline
Uniform (rectangular window)
&
$K(u)=\frac{1}{2}$
, support: $|u| \leq 1$\\
\hline
Cosine
&
$K(u)=\frac{\pi}{4} \cos \left(\frac{\pi}{2} u\right)$
, support: $\left|u_{1}\right| \leq 1$\\
\hline
\end{tabular}
\caption{The basis functions of the different kernels implemented}
\label{table:kernel}
\end{center}
\end{table}

\ref{eq:rosen} is the estimator of $p(x)$ called the Rosenblatt Estimator and it can be rewritten as,

$$\hat{p}_{n}^{R}(x)=\frac{1}{2 n h} \sum_{i=1}^{n} I\left(x-h<X_{i} \leq x+h\right)=\frac{1}{n h} \sum_{i=1}^{n} K_{0}\left(\frac{X_{i}-x}{h}\right),$$

where $n$ is the total number of samples in $D$, $h$ is the bandwidth of the kernel, $K_{0}$ is the selected kernel basis function and $x$ is the point at which the kernel density estimate is to be calculated. Generalizing further the equation can be simplified to

\begin{equation}\label{eq:kernelfinal}
\hat{p}_{n}(x)=\frac{1}{n h} \sum_{i=1}^{n} K\left(\frac{X_{i}-x}{h}\right)
\end{equation}

\ref{eq:kernelfinal} is the kernel density estimator and $\int K(u) d u=1$, where $K$ is called the kernel and $h$ is the bandwidth associated with a particular estimator \cite{silverman1986density}.

\begin{figure}[t!]
\includegraphics[width=\linewidth]{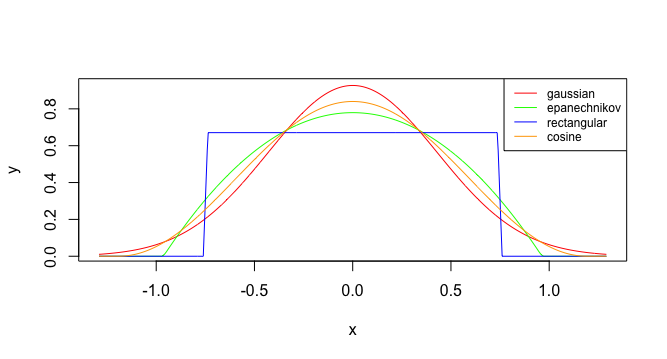}
\caption{All of the implemented kernels on a common coordinate system}
\label{fig: allkern}
\end{figure}

\subsubsection{Multivariate Density Estimation}\label{mde}
This application is concerned with multivariate density estimation. Consider that $X_{1}, \ldots, X_{n}$ constitute an i.i.d. $q$-vector $(X_{i} \in \mathbb{R}^{q}, \text{for some} q > 1)$ having a common PDF $f(x)=f\left(x_{1}, x_{2}, \ldots, x_{q}\right)$. Let $X_{i s}$ denote the $s$th component of $X_{i}(s=1, \ldots, q)$. Using the product kernel function (separability of kernels) constructed from the product of univariate kernel functions, the PDF $f(x)$ is estimated by
$$\hat{f}(x)=\frac{1}{n h_{1} \ldots h_{q}} \sum_{i=1}^{n} K\left(\frac{X_{i}-x}{h}\right)$$
where
$$K\left(\frac{X_{i}-x}{h}\right)=k\left(\frac{X_{i 1}-x_{1}}{h_{1}}\right) \times \cdots \times k\left(\frac{X_{i q}-x_{q}}{h_{q}}\right),$$
and where $k(\cdot)$ is a univariate kernel function.\\

Similar to the univariate case, the optimal smoothing parameter $h_{opt}$ should balance the squared bias and variance term, i.e., $h_{s}^{4}= O\left(\left(n h_{1} \ldots h_{q}\right)^{-1}\right)$ for all $s$. Thus, $h_{s}=c_{s} n^{-1 /(q+4)}$ for some positive constant $c_{s}(s=1, \ldots, q)$.  It is assumed that the data is independently distributed over all $i$. The multivariate kernel density estimator is capable of capturing general dependence among the different components of $X_{i}$ \cite{tsybakov, murphy}.

\subsection{Mean square error of kernel estimators} \label{mse}

A basic measure of accuracy of the estimator is the Mean Square Error (MSE) also called the mean squared risk at any arbitrary point $x_{0} \in \mathbb{R}$. According to \cite{tsybakov}, the MSE is defined as 

\begin{equation}\label{eq:MSE}
\operatorname{MSE}=\operatorname{MSE}\left(x_{0}\right) \triangleq \mathbf{E}_{p}\left[\left(\hat{p}_{n}\left(x_{0}\right)-p\left(x_{0}\right)\right)^{2}\right]
\end{equation}

where $\mathbf{E_{p}}$ denotes the expectation with respect to the random distribution of $X_{1}, X_{2},...,X_{n}$ denoted by $p$

$$\mathbf{E}_{p}\left[\left(\hat{p}_{n}\left(x_{0}\right)-p\left(x_{0}\right)\right)^{2}\right] \triangleq \int \ldots \int\left(\hat{p}_{n}\left(x_{0}, x_{1}, \ldots, x_{n}\right)-p\left(x_{0}\right)\right)^{2} \prod_{i=1}^{n}\left[p\left(x_{i}\right) d x_{i}\right]$$

We have
$$\mathrm{MSE}=b^{2}\left(x_{0}\right)+\sigma^{2}\left(x_{0}\right)$$
where 
$$b\left(x_{0}\right)=\mathbf{E}_{p}\left[\hat{p}_{n}\left(x_{0}\right)\right]-p\left(x_{0}\right)$$
and
$$\sigma^{2}\left(x_{0}\right)=\mathbf{E}_{p}\left[\left(\hat{p}_{n}\left(x_{0}\right)-\mathbf{E}_{p}\left[\hat{p}_{n}\left(x_{0}\right)\right]\right)^{2}\right].$$

The quantities $b(x_{0})$ and $\sigma^{2}\left(x_{0}\right)$ are called the bias and variance of the estimator at a particular point $x_{0}$. As seen in the previous section, the error of an estimator can be decomposed into a sum of its bias and variance at a particular point.

\begin{figure}
\includegraphics[width=\linewidth]{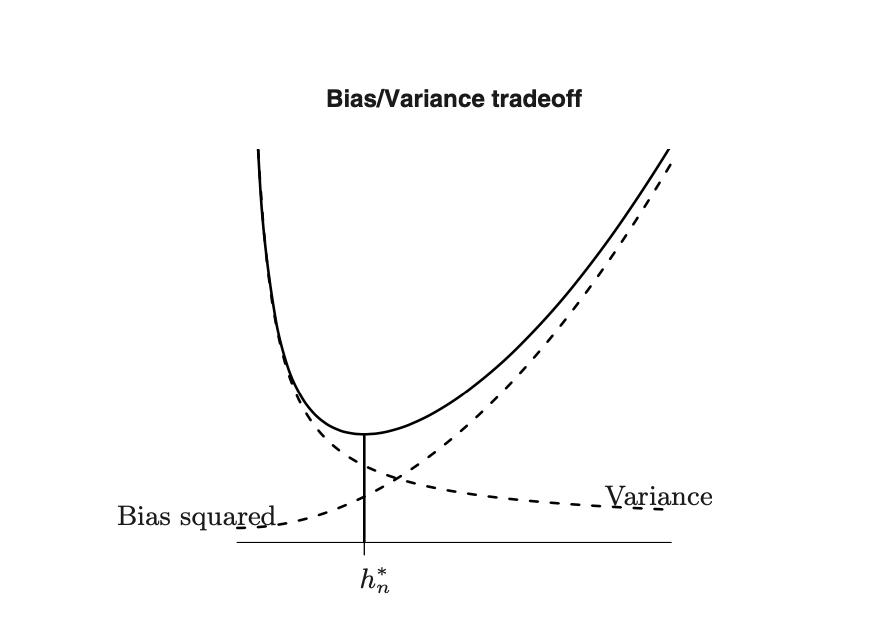}
\caption{Squared bias, variance and mean squared error (solid line) as functions of $h$ as seen in \cite{tsybakov}}
\label{fig: decomp2}
\end{figure}

To better understand how the MSE depends on the bias and variance, each of these terms are analyzed separately as seen in \cite{tsybakov}\\

It is known that the MSE at a random point $x$ is 

$$\begin{aligned} \operatorname{MSE}(\hat{p}_{n}\left(x\right))& \equiv \mathrm{E}\left\{[\hat{p}_{n}\left(x\right)-p(x_{0})]^{2}\right\} \\ &=\operatorname{var}(\hat{p}_{n}\left(x\right))+[\mathrm{E}(\hat{p}_{n}\left(x\right))-p(x_{0})]^{2} \\ & \equiv \operatorname{var}(\hat{p}_{n}\left(x_{0}\right))+[\operatorname{bias}(\hat{p}_{n}\left(x\right))]^{2} \end{aligned}$$\\

To analyze the bias the Taylor Series expansion formula\cite{tsybakov} is used. For an univariate function g$(x)$ that is $m$ times differentiable,

$$\begin{aligned} g(x)=& g\left(x_{0}\right)+g^{(1)}\left(x_{0}\right)\left(x-x_{0}\right)+\frac{1}{2 !} g^{(2)}\left(x_{0}\right)\left(x-x_{0}\right)^{2}+\\ & \cdots+\frac{1}{(m-1) !} g^{(m-1)}\left(x_{0}\right)\left(x-x_{0}\right)^{m-1}+\frac{1}{m !} g^{(m)}(\xi)\left(x-x_{0}\right)^{m} \end{aligned},$$
where $$g^{(s)}\left(x_{0}\right)=\left.\frac{\partial^{s} g(x)}{\partial x^{s}}\right|_{x=x_{0}},$$ and $\xi$ lies between $x$ and $x_{0}.$

The bias term $\text {bias } (\hat{p}_{n}(x))$  can be simplified as

$$=\mathrm{E}\left\{\frac{1}{n h} \sum_{i=1}^{n} k\left(\frac{X_{i}-x}{h}\right)\right\}-p(x)$$
$$=h^{-1} \mathrm{E}\left[k\left(\frac{X_{1}-x}{h}\right)\right]-p(x)$$
$$\text {(by identical distribution)}$$
$$=h^{-1} \int p\left(x_{1}\right) k\left(\frac{x_{1}-x}{h}\right) d x_{1}-p(x)$$
$$=h^{-1} \int p(x+h v) k(v) h d v-p(x)\\ $$
$$(\text{change of variable} :  x_{1}-x = hv)$$
$$=\int\left\{p(x)+p^{(1)}(x) h v+\frac{1}{2} p^{(2)}(x) h^{2} v^{2}+O\left(h^{3}\right)\right\} k(v) d v-p(x) $$
$$=\left\{p(x)+0+\frac{h^{2}}{2} p^{(2)}(x) \int v^{2} k(v) d v+O\left(h^{3}\right)\right\}-p(x)$$
$$=\frac{h^{2}}{2} p^{(2)}(x) \int v^{2} k(v) d v+O\left(h^{3}\right)$$

where the $O\left(h^{3}\right)$ term comes from 

$$(1 / 3 !) h^{3}\left|\int f^{(3)}(\tilde{x}) v^{3} k(v)\right| d v \leq C h^{3} \int\left|v^{3} k(v) d v\right|=O\left(h^{3}\right),$$

where $C$ is a positive constant, and $\tilde{x}$ lies between $x$ and $x+hv$.

It has been assumed that $p(x)$ is three times differentiable, however, one can also assume that it is two times differentiable and subsequently weaken the initial assumption. In that case, bias is written as

$$\begin{aligned} \operatorname{bias}(\hat{p}_{n}(x)) &=\mathrm{E}(\hat{p}_{n}(x))-p(x) \\ &=\frac{h^{2}}{2} p^{(2)}(x) \int v^{2} k(v) d v+o\left(h^{2}\right) \end{aligned}$$\\

Next analyzing the variance term, where $\operatorname{var}(\hat{p}_{n}(x))=$

$$\operatorname{var}\left[\frac{1}{n h} \sum_{i=1}^{n} k\left(\frac{X_{i}-x}{h}\right)\right]$$
$$=\frac{1}{n^{2} h^{2}}\left\{\sum_{i=1}^{n} \operatorname{var}\left[k\left(\frac{X_{i}-x}{h}\right)\right]+0\right\}$$
$$\text{(by independence)}$$

$$=\frac{1}{n h^{2}} \operatorname{var}\left(k\left(\frac{X_{1}-x}{h}\right)\right)$$
$$\text{(by identical distribution)}$$
$$=\frac{1}{n h^{2}}\left\{\mathrm{E}\left[k^{2}\left(\frac{X_{1}-x}{h}\right)\right]-\left[\mathrm{E}\left(k\left(\frac{X_{1}-x}{h}\right)\right)\right]^{2}\right\}$$
$$=\frac{1}{n h^{2}}\left\{\int p\left(x_{1}\right) k^{2}\left(\frac{x_{1}-x}{h}\right) d x_{1}\right.
\left.\quad-\left[\int p\left(x_{1}\right) k\left(\frac{x_{1}-x}{h}\right) d x_{1}\right]^{2}\right\}$$$$=\frac{1}{n h^{2}}\left\{h \int p(x+h v) k^{2}(v) d v\right.\left.-\left[h \int p(x+h v) k(v) d v\right]^{2}\right\}$$
$$=\frac{1}{n h^{2}}\left\{h \int\left[p(x)+p^{(1)}(\xi) h v\right] k^{2}(v) d v-O\left(h^{2}\right)\right\} $$
$$=\frac{1}{n h}\left\{p(x) \int k^{2}(v) d v+O\left(h \int|v| k^{2}(v) d v\right)-O(h)\right\}$$
$$=\frac{1}{n h}\{\kappa f(x)+O(h)\}$$
 
 where,
 
 $$\kappa=\int k^{2}(v) d v$$
 
Thus it is seen that by the correct selection of $h$ the MSE for $p(x)$ can be minimized. By choosing $h= cn^{1/\alpha}$ the conditions for the consistent estimation of $p(x)$ are satisfied. However, the question still remains regarding what values of $c$ and $\alpha$ should be used. For a given sample size $n$, if $h$ is too small the resulting estimator will have a small bias but a large variance. On the other hand if $h$ is too large, the estimator will have a large bias but a small variance. So to minimize $\operatorname{MSE}(\hat{p}_{n}(x))$ the bias and variance terms have to be balanced \cite{tsybakov, shalev}.\\
 
 The optimal choice of $h$ should satisfy 
 
 $$d \operatorname{MSE}(\hat{p}_{n}({x})) / d h=0$$
 
 It can be shown that the value of $h$ which minimizes MSE at a point $x$ is given by 
 
 $$h_{opt} = c(x)n^{-1/5}$$
 
 where 
 
 $$c(x)=\left\{\kappa p(x) /\left[\kappa_{2} p^{(2)}(x)\right]^{2}\right\}^{1 / 5}$$
 
 It is important to note that $\operatorname{MSE}(\hat{p}_{n}({x}))$ is a calculation at a particular point $x$, and that the value of $h$ minimizing the MSE is different depending on the choice of $x$. For instance, the value of $h$ that minimizes the error at a point at the tail end of a distribution is different from the value of $h$ that would minimize MSE if $x$ was the average \cite{tsybakov}. 

The main concern is the minimization of error and subsequently the risk while computing the estimators using kernel density, globally - that is, for all $x$ in the support of $p(x)$. In this case,the optimal $h$ is obtained minimizing the mean integrated square error (MISE) which is discussed in the next section. 

\section{Cross Validation to minimize risk}\label{cv}
This section explains the method by which the optimal bandwidth $h_{opt}$ is determined for the kernel density estimator as seen in \cite{tsybakov, Ala14}. In the previous section, the MSE was defined in \ref{eq:MSE} as

$$\mathrm{MSE}=\operatorname{MSE}\left(x_{0}\right) \triangleq \mathbf{E}_{p}\left[\left(\hat{p}_{n}\left(x_{0}\right)-p\left(x_{0}\right)\right)^{2}\right]$$

which is calculated at a fixed arbitrary point $x_{0}$. However, in analyzing the risk of the estimator, the global risk must be analyzed. The global risk is defined as the Mean Integrated Squared error which can be calculated as follows

\begin{equation}\label{eq:MISE1}
\operatorname{MISE} \triangleq \mathbf{E}_{p} \int\left(\hat{p}_{n}(x)-p(x)\right)^{2} d x
\end{equation}

The MISE can be rewritten as $\mathrm{MISE}(h)$  to denote that it is a function of the bandwidth $h$ and the ideal value of $h$ can be defined as

\begin{equation}
h_{\mathrm{id}}=\arg \min _{h>0} \operatorname{MISE}(h)
\end{equation}

Since $\operatorname{MISE}(h)$ depends on the unknown value of $p$ (which is not known),  a different approach to find the $h_{\mathrm{id}}$ must be employed. In this case, an unbiased estimation of the risk through leave one out cross validation can be used. Instead of minimizing $\operatorname{MISE}(h)$ the approximately unbiased estimator of $\operatorname{MISE}(h)$ is minimized.

It is important to note that
\begin{equation}\label{eq:MISE2}
\operatorname{MISE}(h)=\mathbf{E}_{p} \int(\hat{p}_{n}(x)-p(x))^{2} dx
=\mathbf{E}_{p}\left[\int (\hat{p}_{n}(x))^{2} dx-2 \int \hat{p}_{n}(x) p(x) dx\right]+\int (p(x))^{2} dx
\end{equation}
The last term does not depend on $h$ and therefore can be excluded from the minimization. The first term is the expectation with respect to the distribution of ${X_{1}, X_{2}, X_{3}...,X_{n}}$ denoted by $p$. Therefore, the minimizer for $h_{\mathrm{id}}$ seeks to minimize the term
\begin{equation}\label{eq:risk1}
J(h) \triangleq \mathbf{E}_{p}\left[\int (\hat{p}_{n}(x))^{2} dx-2 \int \hat{p}_{n}(x) p(x) dx\right]
\end{equation}
where $J(h)$ is the risk function. The second term,
\begin{equation}\label{eq:risk2}
\int \hat{p}_{n}(x) p(x) dx
\end{equation}
can be written as $$\mathbf{E_{p}} [\hat{p}(x)],$$ where $\mathbf{E}_{x}$ denotes the expectation with respect to $\mathbf{X}$ and not with respect to the random variables $\mathbf{X}_{i}$. Therefore, this term is the expectation of $\hat{p}_{n}(x)$ with respect to $p(x)$. Rewriting the integral in \ref{eq:risk2} as
\begin{equation}\label{eq:loocv1}
\hat{p}_{n,-i}(x)=\frac{1}{(n-1) h} \sum_{j \neq i} K\left(\frac{X_{j}-x}{h}\right)
\end{equation}
which is the leave one out kernel estimator for $f(X_{i})$ . Finally, the first term in \ref{eq:MISE2}, $\int (\hat{p}_{n}(x))^{2} dx$ can be estimated as
\begin{equation}\label{eq:loocv2}
 \int (\hat{p}_{n}(x))^{2} d x =\frac{1}{n^{2} h^{2}} \sum_{i=1}^{n} \sum_{j=1}^{n} \int k\left(\frac{X_{i}-x}{h}\right) k\left(\frac{X_{j}-x}{h}\right) d x \\ 
 =\frac{1}{n^{2} h} \sum_{i=1}^{n} \sum_{j=1}^{n} \bar{k}\left(\frac{X_{i}-X_{j}}{h}\right)
\end{equation}
where $\bar{k}(v)=\int k(u) k(v-u) d u$ is the twofold convolution kernel derived from $k(\cdot)$. If $$k(v)=\exp \left(-v^{2} / 2\right) / \sqrt{2 \pi},$$ a standard normal kernel, then $$\bar{k}(v)=\exp \left(-v^{2} / 4\right) / \sqrt{4 \pi}$$, a normal kernel (i.e. normal PDF) with mean zero and variance 2, which follows since two independent $N(0,1)$ random variables sum to a $N(0,2)$ random variable\cite{tsybakov}.

For this project, the convolution kernel is computed analytically in the programming language R. Combining \ref{eq:loocv1} and \ref{eq:loocv2}, the leave one out cross validation estimator which minimizes $h$ is obtained as

\begin{equation} \label{eq:loocv3}
C V_{f}(h)=\frac{1}{n^{2} h}  \sum_{i=1}^{n} \sum_{j=1}^{n} \bar{k}\left(\frac{X_{i}-X_{j}}{h}\right) \\- \frac{2}{n(n-1) h} \sum_{i=1}^{n} \sum_{j \neq i, j=1}^{n} k\left(\frac{X_{i}-X_{j}}{h}\right).
\end{equation}

This equation can be easily generalized for multivariate estimation
$$\begin{aligned} C V_{f}\left(h_{1}, \ldots, h_{q}\right)=\frac{1}{n^{2}} & \sum_{i=1}^{n} \sum_{j=1}^{n} \bar{K}_{h}\left(X_{i}, X_{j}\right) -\frac{2}{n(n-1)} \sum_{i=1}^{n} \sum_{j \neq i, j=1}^{n} K_{h}\left(X_{i}, X_{j}\right), \end{aligned}$$
where
$$\begin{aligned} K_{h}\left(X_{i}, X_{j}\right) &=\prod_{s=1}^{q} h_{s}^{-1} k\left(\frac{X_{i s}-X_{j s}}{h_{s}}\right) \\ \bar{K}_{h}\left(X_{i}, X_{j}\right) &=\prod_{s=1}^{q} h_{s}^{-1} \bar{k}\left(\frac{X_{i s}-X_{j s}}{h_{s}}\right), \end{aligned}$$
and $\bar{k}(v)$ is the two fold convolution kernel based upon $k(\cdot)$, where $k(\cdot)$ is the univariate kernel function.

Note that $h_{\mathrm{id}}$ can be found by using different numerical search algorithms. For this particular problem, a range of values for $h$ is defined such that $h = {h_{1}, h_{2}, h_{3}….,h_{n}}$. For every value of $h$ , $CV_{f}(h)$ is computed and then the value of $h$ for which $CV_{f}(h)$ is minimum \cite{tsybakov, Ala14} is obtained. \\

Thus, $CV_{f} (h)$ yields an unbiased estimator of $\operatorname{MISE}(h)$, which is independent of $h$. This means that the functions $MISE(h)$ and $\mathbf{E_{p}} [CV_{f}(h)]$ have the same minimizers. In turn, the minimizers of $\mathbf{E_{p}} [CV_{f}(h)]$ can be approximated by those of the function $CV(\cdot)$ which can be computed from the observations $X_{1},X_{2},X_{3}...,X_{n}$. This means that 
$$h_{CV} = \arg \min CV(h) = h_{\mathrm{id}}.$$

As discussed in this chapter, given a data set of random variables, the main aim is to estimate the density of these points through non-parametric methods by using the kernel density estimator. Four kernel basis functions are chosen: Gaussian, Epanechnikov, uniform (boxcar/rectangular window) and cosine. The best value of $h$ for each of these methods is learned through leave one out cross validation. The chosen value of $h$ minimizes the loss function and subsequently the global risk of the chosen estimation method. The mathematics of the exact process employed is discussed in the next chapter.

%% file: chapter3.tex
\chapter{ALGORITHM AUTOMATION}
\label{chap:automation}
In this chapter, the problem statement and the specific application of the concepts and methods discussed in the previous chapter are discussed. While Chapter 2 provided a general overview of the proposed method and concepts employed, this chapter analyzes the specific applications of the concepts previously discussed to the chosen data set. Referring to the mathematical model presented in \cite{bahman} this chapter seeks to relate the theoretical concepts to the practical implementation.\\

Following the method presented in \cite{bahman}, the ECG data is obtained from the Preterm Infant Cardio-Respiratory Signals (PICS) database \cite{database, PhysioNet}. This database contains simultaneous ECG and respiration recordings of ten preterm infants collected from the Neonatal Intensive Care Unit (NICU) of the University of Massachusetts Memorial Healthcare. Statistical features based on linear estimates of heart rate are used to predict episodes of bradycardia. 

\section{Data Collection}

As presented in \cite{bahman}, ten pre-term infants were studied, with post-conceptional age of 29 3/7 to 34 2/7 weeks (mean: 31 1/7 weeks) and study weights of 843 to 2100 grams (mean: 1468 grams). The infants were spontaneously breathing room air and did not have any congenital or perinatal infections or health complications. A single channel of a 3-lead electrocardiogram (ECG) signal was recorded at 500 Hz for ~20-70 hours per infant. In absence of an ECG channel, a compound ECG signal was recorded (250Hz) \cite{database}. 

Each ECG signal is pre-processed and segmented such that each segment contains both normal and unhealthy beats. R-peak information is extracted using the Pan-Tompkin algorithm\cite{pantomp} and estimate the non specific probability density function. After setting a desired level of false alarm to be tolerated by the system, a threshold region is found using the estimated density and a generated threshold plane. R-peaks from further along in time are used to test against this threshold region to determine the onset of near-term bradycardia \cite{bahman}.

\section{Non-parametric prediction}

Once the R-peaks have been obtained, a non-parametric method is employed to determine the kernel-based probability density function which is used to predict the onset of bradycardia in the future \cite{tsybakov, bahman}. 

Assuming the number of R peaks in an ECG segment is $N$, the R-tuple is defined as $x=(t_{n},R_{n})$, n=1,...,N and $x_{n} \in \mathcal{X}$, where $t_{n}$ is the sample at which the peak occurs and $R_{n}$ is the amplitude of the peak. Therefore, the set of R tuples $X= {x_{1}, x_{2},....x_{N}}$ is i.i.d. and drawn from an unknown distribution $p(x)$. As discussed in Chapter 2, using the kernel density estimator, one can estimate $p(x)$ as
$$\hat{p}_{H}(x)=\frac{1}{N} \sum_{n=1}^{N} K_{H}\left(x-x_{n}\right)$$
for some positive definite bandwidth matrix H. without loss of generality, it is assumed that $H=h^{2} I_{2}$ , where $I_{2}$ is the (2x2) identity matrix. The best value of $h$ is learned using leave one out cross validation to ensure unbiased estimation \cite{bahman, Ala14}.

In order to construct the hypothesis set, a probability of false alarm $P_{FA}$ is declared such that the hypothesis testing produces $(1-P_{FA})$ confidence. For this application, $P_{FA}$ = $5\%$ and therefore the hypothesis testing produces $95\%$ confidence with which the future onset of bradycardia is predicted. The null hypothesis is defined as $\mathcal{H}_{0}$ as the hypothesis that the density of the next R-tuple $x_{N+1}$ is the same as that of the previous R-tuples in the set $\mathcal{X}_{N}$; that is, $\mathcal{H}_{0}: x_{N +1} = x$ ,for all possible values of $x \in \mathcal{X}$ \cite {bahman}. The aim is to construct a confidence set $\mathcal{A}_{\mathcal{X}}$ , that consists of all values $X_{N}$, such that the probability of the next R-tuple, $x_{N+1}$, belonging to this set satisfies the following condition 
$$\operatorname{Pr}\left(x_{N+1} \in \mathcal{A}_{\mathcal{X}}\right) \geq\left(1-P_{\mathrm{FA}}\right)$$
 The kernel density estimator $\hat{p}_{H}^{a}(x)$ based on the augmented data set $X_{N} \cup\{x\}$ for a fixed value of $x \in \mathcal{X}$. is then used The rank of $\hat{p}_{H}^{a}\left(x_{1}\right), \ldots, \hat{p}_{H}^{a}\left(x_{N+1}\right)$ is uniformly distributed under the null hypothesis. Thus, for each value of $x$, the p-value $\eta_{x}$ is given by,
$$\eta_{x}=\frac{1}{N+1} \sum_{n=1}^{N} \mathbb{I}\left(\hat{p}_{H}^{a}\left(x_{n}\right) \leq \hat{p}_{H}^{a}(x)\right),$$
where $\mathbb{I}$ is the indicator function. Therefore, the confidence set $(1-P_{FA})$ is defined as 
$$\mathcal{A}_{\mathcal{X}}=\left\{x: \eta_{x} \geq P_{\mathrm{FA}}\right\}$$

$\mathcal{A}_{\mathcal{X}}$ is distribution free \cite{lei2013distribution} and only determined using the finite set $X_{N}$\cite{shafer2008tutorial}. This method is impractical since $X_{N}$ are merely limited observations drawn from the original unknown distribution. Therefore,  a larger set which contains $\mathcal{A}_{\mathcal{X}}$ and can be easily constructed named $\mathcal{B}_{\mathcal{X}}$ is defined. This set is easier to compute and preserves accuracy \cite{bahman}.

Additionally, $y_{n}=\hat{p}_{H}\left(x_{n}\right)$ for $n=1,....,N$ is also defined and it is assumed that $y_{i}$ is sorted in ascending order, that is $y_{1} \leq \cdots \leq y_{N}$. The prediction set $\mathcal{B}_{\mathcal{X}}$ can be constructed as
\begin{equation}\label{eq:confidenceset}
\mathcal{B}_{\mathcal{X}}=\left\{x: \hat{p}_{H}(x) \geq \mathcal{C}_{k}\right\},
\end{equation}
The threshold plane $\mathcal{C}_{k}$ can be compute as 
$$\mathcal{C}_{k} = y_{k} - \left(K_{H}(0) / N|H|^{1 / 2}\right)$$
where $k=[(N+1) P_{FA}]$.

The set $\mathcal{B}_{\mathcal{X}}\supset \mathcal{A} x$ satisfies 
$$\operatorname{Pr}\left(x_{N+1} \in \mathcal{A}_{\mathcal{X}}\right) \geq\left(1-P_{\mathrm{FA}}\right) \Longrightarrow \operatorname{Pr}\left(x_{N+1} \in \mathcal{B}_{\mathcal{X}}\right) \geq\left(1-P_{\mathrm{FA}}\right).$$

In other words, the prediction set $\mathcal{B}_{\mathcal{X}}$ is the projection of the estimated density which is above the threshold $\mathcal{C}_{k}$ \cite{bahman}. Once this prediction set has been obtained, any R-tuple $x_{m}=\left(t_{m}, R_{m}\right)$ is predicted with $95\%$ confidence to be the onset of bradycardia if 
\begin{equation}\label{teststat}
x_{m} \notin \mathcal{B}_{\mathcal{X}}, m> N.
\end{equation}
The proposed model employs unsupervised learning since the training data is unlabeled and predictions are made on all unseen points. As with all machine learning approaches, the data is segmented into training, validation and testing sets. However, before this can be done, the ECG signals must be analyzed and preprocessed. 

\section{Data Preprocessing}
Each ECG signal is preprocessed by first removing baseline wander, which is a low frequency of around 0.5-0.6Hz. It is removed by using a high pass filter with cut off frequency between 0.5 and 0.6Hz. There are two other values present in the ECG header files obtained from the database:
\begin{itemize}
\item
$\emph{Gain}$: This is a floating-point number that specifies
the difference in sample values that would be observed if a step of one physical unit
occurred in the original analog signal. For ECGs, the gain is usually roughly equal to
the R-wave amplitude in a lead that is roughly parallel to the mean cardiac electrical
axis. If the gain is zero or missing, this indicates that the signal amplitude is not
calibrated. For this dataset, every ECG recording has its own value of gain \cite{database}.
\item
$\emph{Base}$: The base is a floating point number that specifies the counter value corresponding to sample 0. For this database, the base counter is 16 which means that sample 0 is shifted to the 16th position in the files \cite{database}.
\end{itemize}

The ECG signals are adjusted for the gain and base values for each file. These values can be obtained from the header file for each infant contained in the database. 

Once the baseline wander has been removed, each signal is segmented into events. The number of events is equal to the number of bradycardia events for that infant which is obtained from the annotation file. For instance, the annotation file for infant 5 contains the onset of 72 bradycardia events. Therefore the cleaned ECG signal for this infant is segmented into 72 events where each event contains some samples before and after the bradycardia onset. For this project, 5000 samples before and 2500 samples after the onset of bradycardia were chosen randomly to be included in each event. Therefore, every event has 7501 samples. 

\begin{table}
\resizebox{\columnwidth}{!}{
\begin{tabular}{ c c c c c c c c c c c}
\hline \hline
Parameters & Infant 1 & Infant 2 & Infant 3 &Infant 4 &Infant 5 &Infant 6 &Infant 7 &Infant 8 &Infant 9 &Infant 10  \\ 
\hline \hline
Bradycardia events & 77 & 72 & 80 & 66 & 72 & 56 & 34 & 28 &97 & 40\\
Duration (hours) & 45.6 & 43.8 & 43.7 & 46.8 & 48.8 & 48.6 & 20.3 & 24.6 & 70.3 & 45.1\\
\hline\hline
\end{tabular}
}
\caption{Number of bradycardia events and duration of ECG for each infant as obtained from \cite{database}}
\label{table:ecg}
\end{table}

\section{Peak detection}

The R-peaks are detected by running the Pan-Tompkin algorithm on each event. Each detected peak $x_{n}$ is a tuple $(t_{n},R_{n})$ for every event for a particular infant. These peaks are used as the data to run the rest of the model for bradycardia prediction. The R peaks are stored in .csv files and are used to generate the training, validation and testing set for the algorithm.

\section{Data set segmentation}

Once the peaks are detected, the entire data set of peaks is randomly shuffled and then segmented into training, validation and testing set according to some pre-determined demarcations. To this end, a percentage of total points to be allocated to each set is determined. The sets are generated using MATLAB following which the training set is used for kernel density estimation. The output of the kernel estimator is used to calculate $C_{k}$ and to generate the confidence set $\mathcal{B}_{x}$ according to \ref{eq:confidenceset} and \ref{teststat}.

\subsection{Training Set}

The training set is used to train the model - that is, this is the set used for kernel density estimation of the original probability density function (pdf) \cite{salev}. The training set contains the location and amplitude of R peaks as detected by the Pan Tompkin algorithm. Before the set can be used for kernel density estimation, the location of R peaks is scaled by the sampling frequency and normalized. The training set is also used to generate the points on the $x$ and $y$ axis (evaluation points) where the kernel density estimation is to be calculated. The results from the estimation are used to compute the threshold $\mathcal{C}_{k}$ and to generate the confidence set $\mathcal{B}_{x}$. 

\subsection{Validation Set}

Before the estimator can be run on the training set, the leave one out cross validation is used on the validation set to learn the best value of $h$. The validation set is used to prevent reuse of data points for both training and cross validation and subsequently prevent bias. The leave one out cross validation is run for the chosen  kernel \cite{bahman}. 

\subsection{Test Set}

The testing condition based on the prediction set  is defined according to \ref{eq:confidenceset}. A point $m$ from the test set is said to be the onset of bradycardia if
\begin{equation}\label{eq:confcond}
x_{m} \notin \mathcal{B}_{x}, m>N.
\end{equation}

The condition $m>N$ is maintained since it is assumed that the points in the test set are separate from the training data (which contains $N$ points).

\subsection{Monte-Carlo simulations}

To obtain the average error Monte-Carlo simulations for different demarcations of the entire event data set. For instance, the data set is split into: a) $60\%,20\%$,$20\%$, b) $70\%,20\%, 10\%$ and c) $70\%, 10\%, 20\%$ to form three separate training, validation and testing sets respectively.

The annotation file of every infant contains the onset of bradycardia while the Pan-Tompkin output contains the location of the peaks. It is reasonably assumed that the peak locations that occur after the bradycardia onset are the bradycardia peaks \cite{database}. The question then arises how far into the future from the onset of bradycardia should the algorithm search to locate the bradycardia peak. It is assumed that it needs to search $k$ samples after the bradycardia onset. Since each event was generated by taking 2500 samples after the onset $m$, $k<2500$ otherwise the end of each event set would be reached

In theory, $m < k < m+2500$. Therefore, to calculate the appropriate value of $k$,a random number between $u$ is generated such that $u \in (0,1]$. The value of $k$ is then calculated as
$$m < k < m + \mathrm{ceil} (u*1500).$$

For a test point, a peak $x_{n}$ is said to be a bradycardia peak if it satisfies the condition for $k$ outlined above. The estimated prediction error (EPE) is calculated as 
$$\mathrm{EPE}=\frac{\text{Number of False Alarms}}{\text{Number of R-tuples tested}}.$$
Since a lower value of EPE indicates that lesser false alarms are generated, the test error is a measure of performance of this method \cite{bahman}. For a fixed value of $P_{FA}$ the total test error for each infant is calculated in Chapter 4.

\subsection{Testing mechanism}

Once $\mathcal{B}_{x}$ has been generated using $\mathcal{C}_{k}$, the convex hull of $\mathcal{B}_{x}$ is calculated to determine its boundary. Every point from the test set is then used to assess whether it  belongs to $\mathcal{B}_{x}$ by determining its position with respect to the boundary. If a point lies outside the boundary, it is said to lie outside $\mathcal{B}_{x}$ and therefore be the onset of bradycardia \cite{bahman}.

The implementation of the test statistic as seen in \ref{teststat} depends on the creation of the \ref{eq:confidenceset}. Starting from the raw ECG signal, the signal is pre-processed, segmented into events and the R-peaks are detected using the Pan-Tompkin algorithm. The peaks are then stored as tuples and used to create the training set on which the kernel density estimator is run. The validation set is  used to estimate $h$ and finally the test error is calculated by evaluating the performance of the algorithm on the test set. This is elaborated in more detail in the Chapter 4.

%% file: chapter4.tex
\chapter{EXPERIMENTS}

\label{chap:experiments}

In this section, the simulation results of the process discussed in Chapter 3 are presented. First each ECG signal is pre-processed by accounting for the gain of the recording machine and the baseline wander. Once this is done, each ECG signal is segmented into events. Each event contains $N$ peaks which are detected using the Pan-Tompkin algorithm. The simulation results are categorized infant to infant - i.e., all the steps of the algorithm are executed for the events generated by each infant and calculate an average test error by running Monte Carlo simulations for each infant. Therefore, for a particular infant, if there are $M$ total events, the Pan-Tompkin is run $M$ times to detect the peaks in each event. The output of the Pan-Tompkin is stored in a large matrix and then shuffled randomly to generate the training, validation and test sets. 

The output of the Pan-Tompkin is a tuple $(t_{n}, r_{n})$ where $t_{n}$ is the location of the peak and $r_{n}$ (where $n=1,2,\dots,N$) is the amplitude at that location \cite{pantomp}. The set of tuples that constitute the training set is the input to the kernel density estimator where each tuple represents an instance of a random variable \cite{bahman}. It is assumed that the tuples are independent and identically drawn from the same underlying distribution. In the first step of the automation the aim is to estimate this distribution using the kernel density estimator. The initial kernel density estimate is calculated using the built in $\emph{density()}$ function in R. However, in calculation of the threshold $\mathcal{C}_{k}$ and creation of the confidence set $\mathcal{B}_{x}$ a custom function designed for this purpose called $\emph{kerneldensity()}$ is used.

\begin{table}[t!]
\begin{center}
\begin{tabular}{|c c c c|}
\hline
Infant & $F_{s}$ (Hz) & Base & Gain  \\ 
 \hline
1 & 250 & 16 & 800.6597\\
2 & 500 & 16 & 1220.7707\\
3 & 250 & 16 & 1140.7954\\
4 & 500 & 16 & 834.3036\\
5 & 500 & 16 & 800.6597\\
6 & 500 & 16 & 800.6141\\
7 & 500 & 16 & 1283.8528\\
8 & 500 & 16 & 1420.7631\\
9 & 500 & 16 & 800.6597\\
10 & 500 & 16 & 800.4159\\
\hline
\end{tabular}
\end{center}
\caption{Information obtained from the header file(.hea) for each infant from the database. }
\label{table:infantinfo}
\end{table}

\begin{figure}[h!]
\centering
\includegraphics[width = \textwidth]{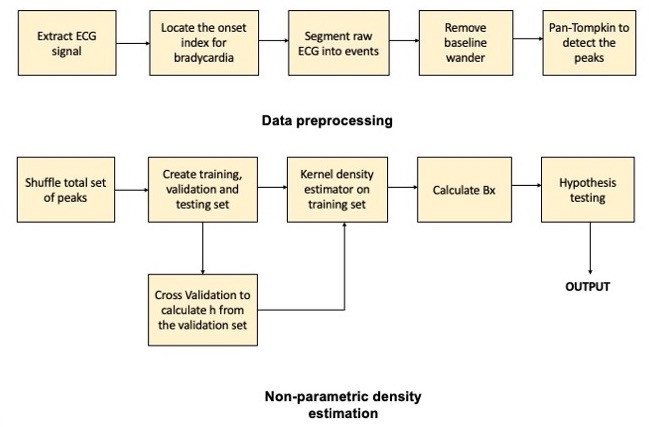}
\caption{Block diagram representation of the preprocessing and density estimation stage. The bradycardia onset indices can be found from the .atr file available in the database used.}
\label{fig: blockdiagram}
\end{figure}

Once the baseline wander has been removed,the peaks and their locations are detected using the Pan-Tompkin algorithm. Following which they are randomly shuffled and used to create the training, validation and testing sets according to some predetermined ratio. Then the training set is used as input to the kernel density estimator. 

The estimator output is used to calculate the threshold $\mathcal{C}_{k}$ and to create the confidence set $\mathcal{B}_{x}$ according to the following equation 
$$\mathcal{B}_{x}=\left\{x: \hat{p}_{H}(x) \geq \mathcal{C}_{k}\right\}.$$
This means that all of the values of $\hat{p}_{H}(x) < \mathcal{C}_{k}$ are rejected. This gives us the desired confidence set.T the convex hull of this shape is calculated to establish its boundary. This is important because for a point $x_{m}$ to be considered the onset of bradycardia it has to satisfy the following equation
$$x_{m} \notin \mathcal{B}_{\mathcal{X}}, m>N.$$
Therefore, if the points in the generated test set lie within the convex hull of $\mathcal{B}_{x}$, they are not considered to be the onset of bradycardia whereas all points lying outside $\mathcal{B}_{x}$ are considered to be the opposite. \\

\begin{figure}[h!]
     \begin{center}
         \includegraphics[width=\linewidth]{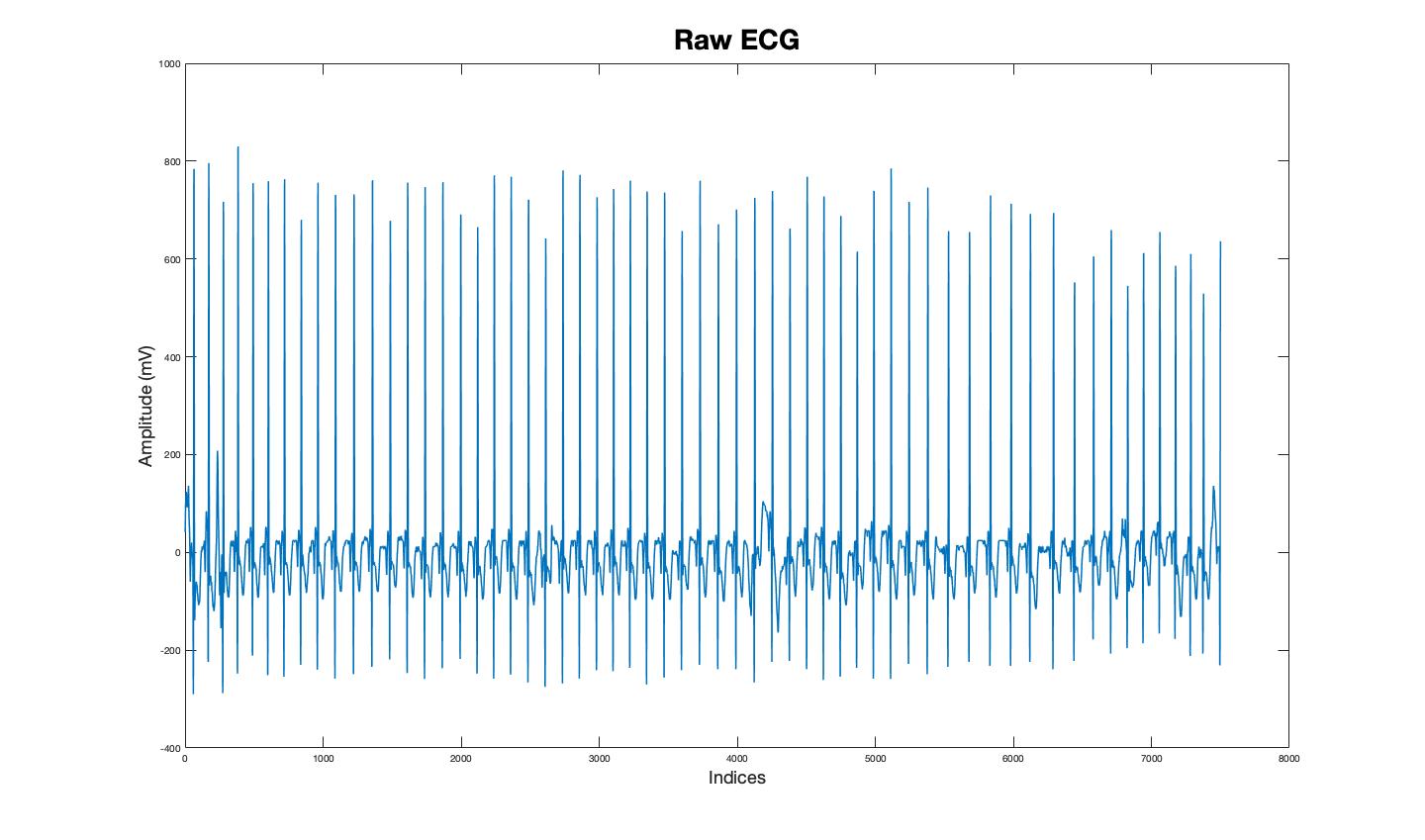}\\
         \includegraphics[width=0.9\linewidth]{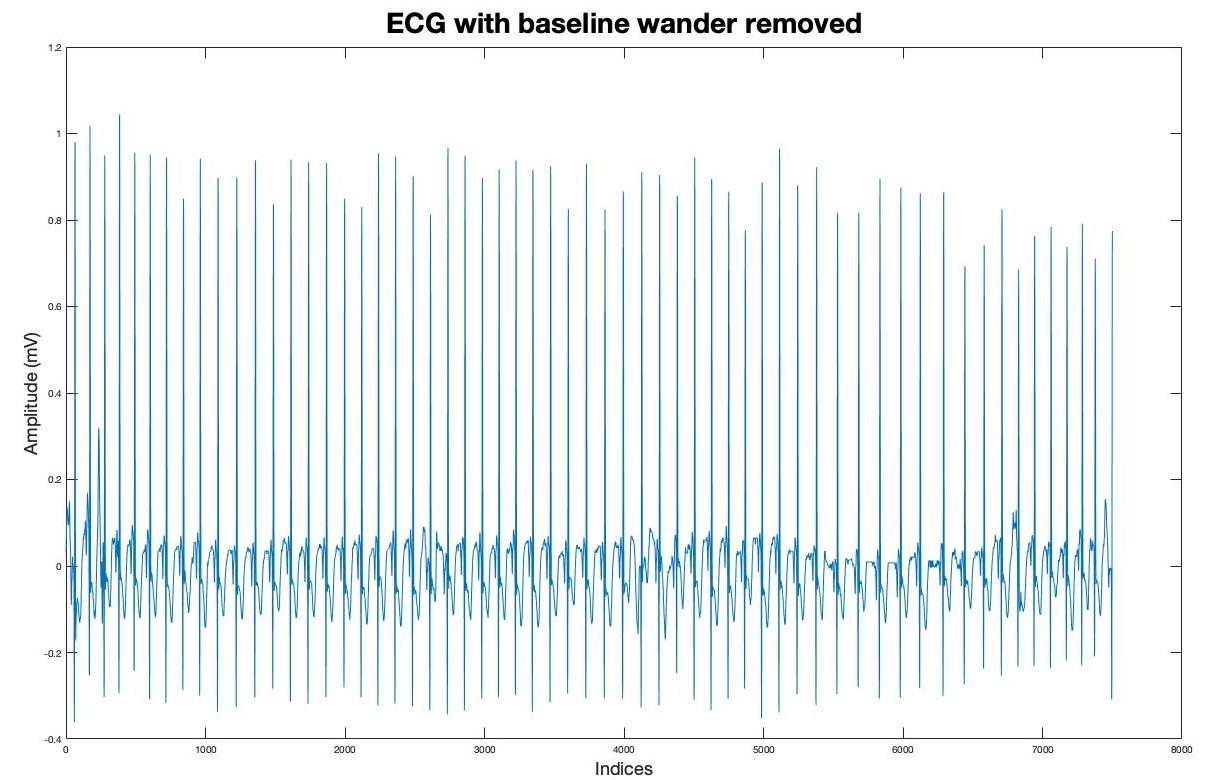}
         \label{fig:rawsig}
     \end{center}
        \caption{A segment of the raw ECG for infant 5, obtained using the $rdsamp()$ command \cite{database}}
\end{figure}

\section{Evaluation Metrics}\label{evalmet}

To evaluate the performance of the kernel density estimator, the following metrics are used \cite{kuhn}:
\begin{itemize}
\item
$\emph{Test error}:$ The test error is defined as the ratio of the number of false classifications and the number of tuples tested. 
\item
$\emph{Sensitivity(Recall)}:$ Sensitivity is defined as the ability of an algorithm to predict a positive outcome when the actual outcome is positive. In this case, it is the ability of the algorithm to predict the onset of bradycardia correctly.
\item
$\emph{Specificity (Precision)}:$ The ability of an algorithm to not predict a positive outcome when the outcome is not positive. In this case, specificity is the ability of the algorithm to classify the healthy heartbeat correctly. 
\item
$\emph{False Discovery Rate}:$ The ratio of all false positive classifications to the total number of all positive classifications. As the name indicates this metric points to all discoveries that are classified as positive but are actually negative.
\item
$\emph{False Omission Rate}:$ As the name suggests, this score is the ratio of all false negative classifications to the sum of all negative classifications. This score displays how many positive instances are falsely omitted as being negative.
\item
$\emph{Accuracy}:$ The ratio of correctly predicted observations to the total observations. This metric is an evaluation of how accurately the algorithm predicts the desired outcome.
\item
$\emph{F1 score}:$ The weighted average of Precision and Recall. This score takes both false positives and false negatives into account. 
\end{itemize}

\begin{figure}[t]
\includegraphics[width=\linewidth]{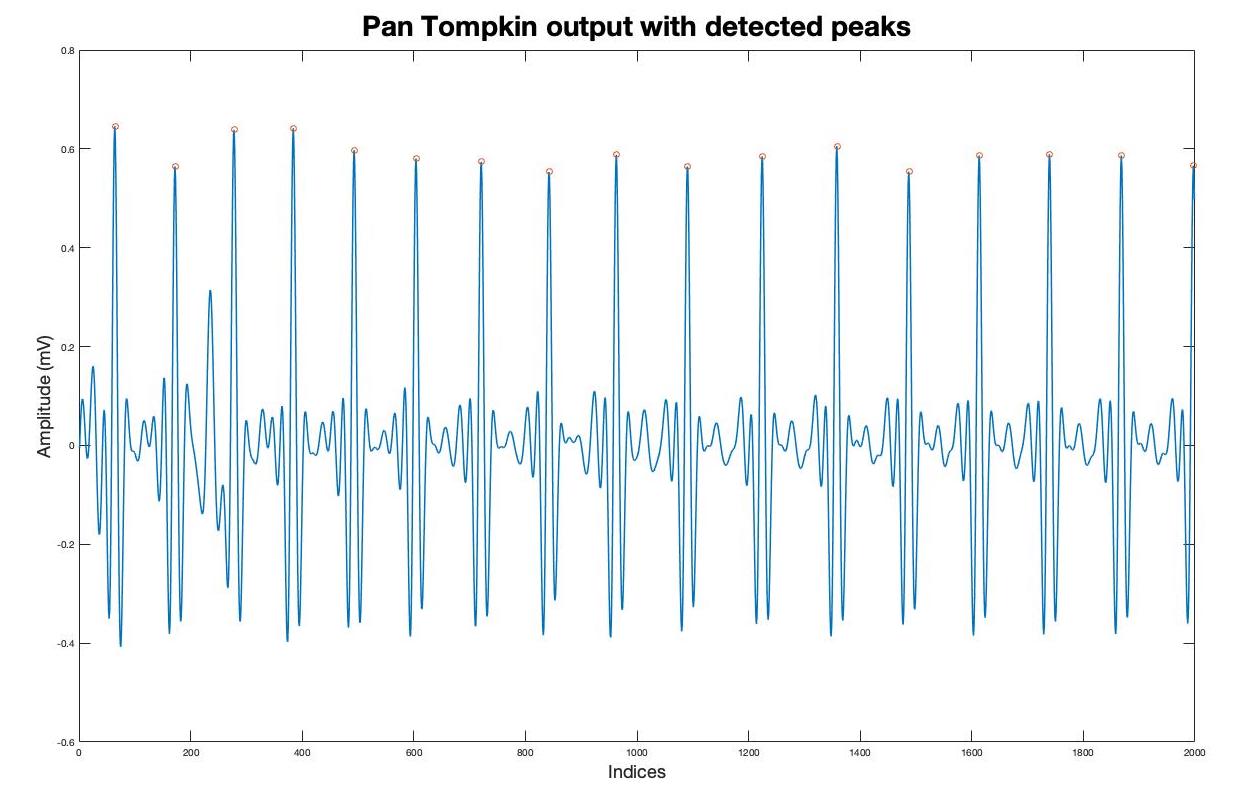}\\
\caption{Pan Tompkin output for an ECG segment for infant 5}
\label{fig: ptout}
\end{figure}

These metrics can be decided from the confusion matrix as seen in \ref{table:confusionmat} and \ref{table:nfant4conf1}. The confusion matrix is the tabular representation of each combination of prediction and actual value.Confusion matrices require a binary outcome. In this case, the positive case is that the point is the onset of bradycardia i.e., it lies outside the convex hull (denoted by 0) and it is not the onset (denoted by 1).  The average test error for infant 5 is calculated in \ref{table:avgtesterror}.
\begin{figure}[thb!]
\includegraphics[width=\linewidth]{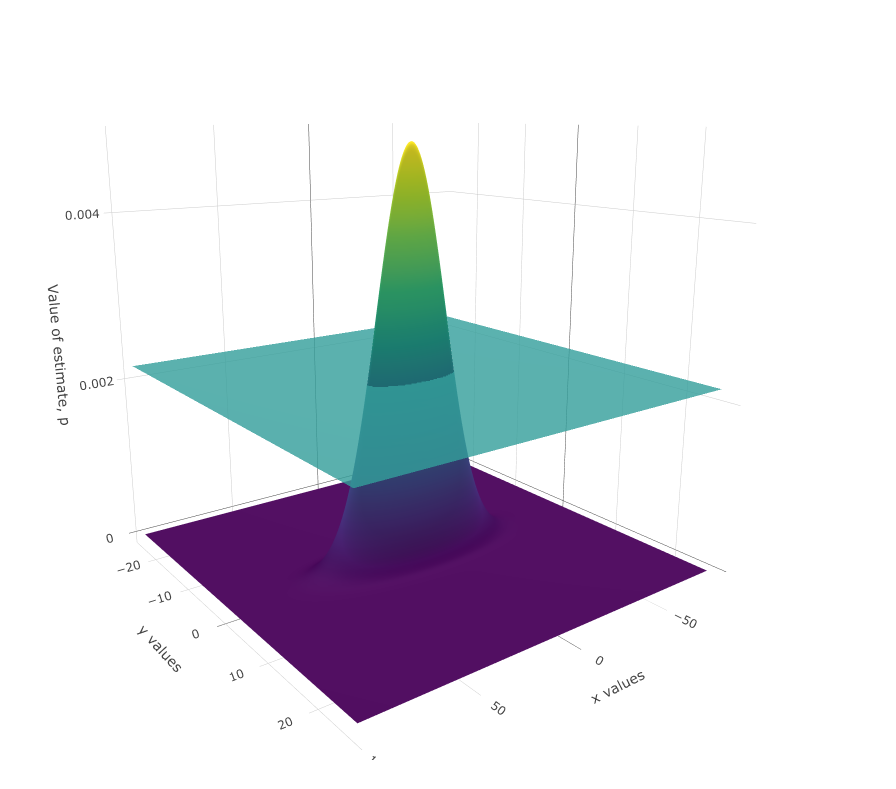}\\
         \caption{Estimate $\hat{p}_{H}(x)$ for Gaussian kernel. }
         \label{fig:gausskern1}
\end{figure}

In \ref{table:avgtesterror}, \textbf{Error1} = average test error calculated when data is split into the training/validation/testing set in the ratio of $.7/.1/.2$. For \textbf{Error2} the split is $.7/.2/.1$ and for \textbf{Error3} the split is $.6/.2/.2$. This error is the result of Monte Carlo simulations run on the data. 20 such simulations are run to calculate the average test error. 
\begin{figure}[thb!]
\includegraphics[width=\linewidth]{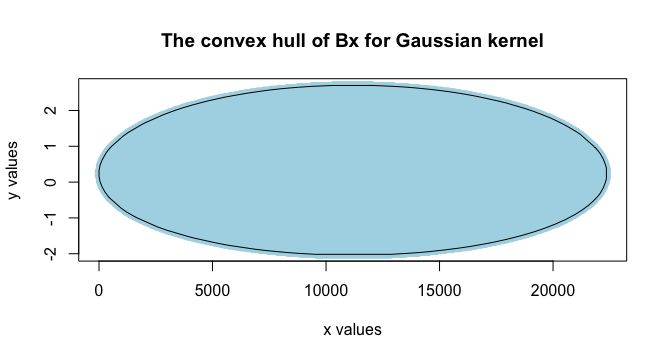}\\
        \caption{Plots showing the kernel density estimate for infant 5 using the gaussian kernel with $h =4.981171$ and $\mathcal{C}_{k} = 0.002179295$ and the obtained confidence set $\mathcal{B}_{x}$. The convex hull for the confidence set is outlined in black.}
        \label{fig:kerngraphs}
\end{figure}
\begin{table}[t]
\begin{center}
\begin{tabular}{|c c c|}
\hline
 Classifications & $\mathbf{Actually Positive}$ & $\mathbf{Actually Negative}$ \\ 
 \hline
$\mathbf{Predicted Positive}$ & True Positives (TP) & False Positives (FP)\\
$\mathbf{Predicted Negative}$ & False Negatives (FN) & True Negatives (TN)\\
\hline
\end{tabular}
\end{center}
\caption{A general confusion matrix}
\label{table:confusionmat}
\end{table}

\begin{table}[t]
\begin{center}
\begin{tabular}{|c c c|}
\hline
 Classifications & \textbf{Actually Positive} & \textbf{Actually Negative} \\ 
 \hline
\textbf{Predicted Positive} & 18 & 26\\
\textbf{Predicted Negative} & 9 & 415\\
\hline
\end{tabular}
\end{center}
\caption{Confusion matrix for Gaussian kernel run on infant 5}
\label{table:nfant4conf1}
\end{table}

\begin{table}[h!]
\begin{center}
\begin{tabular}{| c | c | c | c | c | c |}
\hline
\textbf{Kernel } & $\mathbf{h}$ & $\mathbf{\mathcal{C}_{k}}$ & \textbf{Error1} & \textbf{Error2} & \textbf{Error3}\\ 
\hline
Gaussian & 4.981171  & 0.002179295 & 7.90000005 & 7.6266667 & 7.95026705\\
Epanechnikov & 10.52929 & 1.02x$10^{-05}$ & 6.9500001 & 6.5284985 & 6.8285046\\
Cosine & 10.23305 & 1.26x$10^{-07}$ & 7.4 & 7.2599999 & 7.4782377\\
Uniform & 2.675163 & 0.004355385 & 8.4401727 & 8.19666675 & 8.45300395\\
\hline
\end{tabular}
\end{center}
\caption{Average test error for each kernel as calculated for infant 5 through Monte Carlo simulation}
\label{table:avgtesterror}
\end{table}
On the other hand, \ref{table:epe} is the result of running the Gaussian kernel on the data only once. The total number of tuples in the test set for infant 5= 468. Therefore, the estimated predictive error (EPE) is calculated as:
$$\mathrm{EPE}=\frac{\text{Number of False Alarms}}{\text{Number of R-tuples tested}}$$
where, Number of False classifications = FP + FN . From \ref{table:confusionmat}, FN = 9 and FP = 26, therefore, 
$$\mathrm{EPE}=\frac{(26+9)}{468} = 7.478\%. $$
As observed this value is very close to the average test error for Gaussian kernels presented in \ref{table:avgtesterror}. 

The value of EPE for other kernels is shown in \ref{table:epe}.  These findings are consistent with the average test error for each kernel presented in \ref{table:avgtesterror}. To evaluate the performance of this algorithm, additional metrics are calculated from \ref{table:confusionmat}.

\begin{table}[h!]
\begin{center}
\begin{tabular}{|c c c|}
\hline
 \textbf{Kernel} & \textbf{Number of points tested} & \textbf{Test error} \\ 
 \hline
Epanechnikov & 468 & $6.41025\%$\\
Cosine &  468 & $7.2649\%$\\
Uniform & 468 & $8.5470\%$\\
\hline
\end{tabular}
\end{center}
\caption{Calculated EPE for Epanechnikov, cosine and uniform kernel run on infant 5}
\label{table:epe}
\end{table}

Finally, all metrics mentioned in \ref{evalmet} are evaluated for the Gaussian Kernel and presented in \ref{table:metrics}.

\begin{table}[h!]
\begin{center}
\begin{tabular}{|c c c|}
\hline
 \textbf{Metric} & \textbf{Formula} & \textbf{Value} \\ 
 \hline
Sensitivity & $\frac{TP}{TP+FN}$ & $0.75$\\
Precision &  $\frac{TP}{TP+FP}$ & $0.4091$\\
False Discovery Rate & $\frac{FP}{TP+FP}$ & $0.5909$\\
False Omission Rate &$\frac{FN}{FN+TN}$ & $0.0212$\\
Accuracy & $\frac{TP+TN}{TP+TN+FP+FN}$ & $0.9252$\\
$\text{F}_{1}$ score & $\frac{TP}{TP+0.5(FP+FN)}$&  $0.5070$\\
\hline
\end{tabular}
\end{center}
\caption{Calculated metrics for infant 5}
\label{table:metrics}
\end{table}

\section{Discussion}

From the results of the Monte Carlo simulations in \ref{table:avgtesterror}, it is observed that the Epanechnikov kernel performs the best with respect to the EPE. This is because the Epanechnikov kernel is optimal with respect to MSE. Due to the MSE minimization employed in this estimation process, it ensures the best performance of the Epanechnikov kernel. On the contrary, the uniform kernel performs the worst out of the four kernels studied. This makes sense since the uniform kernel assigns equal weight to all points in its support. 

These findings are consistent with \ref{table:epe} where the Epanechnikov kernel performs the best and the uniform kernel performs the worst in terms of the test error. The accuracy of the method is 92.52\% as seen in \ref{table:metrics}. The code for this implementation can be found in \cite{Sinjini2020}.

%% file: chapter5.tex
\chapter{FUTURE WORKS}
\label{chap:future works}
\acresetall

The previous sections have highlighted the motivations for the work presented in this document, the related theories implemented and the results obtained from this process. As seen in Chapter 4, the algorithm is evaluated using various performance metrics and it is important to note that the average testing error calculated is slightly higher than the proposed $P_{FA} = 5\%$ in \cite{bahman}. 

 The method of non-parametric density estimation can also be extended to other areas of health care such as early detection autism, MRI imaging, early detection of cancer, study of cardio-respiratory diseases in adults etc. In addition to advances in nonparametric modeling, Bayesian nonparametric modeling has found many applications in the field \cite{moraffah2020bayesian, ghosh2003bayesian}.  
 
 The results obtained can also be used to study the effects of early detection of bradycardia in preterm infants. Outside of the health care industry - multiple object tracking using Bayesian modeling \cite{moraffah2019BNP, moraffah2020bayesian}, non-parametric bayesian estimation of noise density for non-linear filters \cite{bib14}, mining of spatio-temporal behaviour on social media using non-parametric modeling \cite{bib15}; are just a few applications where the discussed method can be extended. In short, the method presented can be used for any applications that employ non-parametric density estimation and the calculation of a Type I error. 